\documentclass[twocolumn,aps]{revtex4}

\usepackage{graphicx}
\usepackage{amssymb}
\usepackage{amsmath}
\usepackage{subfigure}
\usepackage{graphpap}
\usepackage{dcolumn}
\usepackage{bm}
\usepackage{color,calc}
\usepackage{longtable}
\usepackage{hyperref}

\usepackage[T1]{fontenc}
\usepackage{lmodern}

\newcommand{\bra}{\langle}
\newcommand{\ket}{\rangle}

\renewcommand{\i}{\text{i}}
\newcommand{\f}{\text{f}}
\renewcommand{\d}{\text{d}}

\begin{document}   
\preprint{}
\title{Gaussian continuum basis functions for calculating high-harmonic generation spectra}
\author{Emanuele Coccia$^{1}$}
\author{Bastien Mussard$^{1,2}$}
\author{Marie Labeye$^3$}
\author{J\'er\'emie Caillat$^3$}
\author{Richard Ta\"{\i}eb$^3$}
\author{Julien Toulouse$^{1}$}
\author{Eleonora Luppi$^{1}$}
\affiliation{
$^1$Sorbonne Universit\'es, UPMC Univ Paris 06, CNRS, Laboratoire de Chimie Th\'eorique, F-75005 Paris, France\\
$^2$Sorbonne Universit\'es, UPMC Univ Paris 06, Institut du Calcul et de la Simulation, F-75005, Paris, France\\
$^3$Sorbonne Universit\'es, UPMC Univ Paris 06, CNRS, Laboratoire de Chimie Physique-Mati\`ere et Rayonnement, F-75005 Paris, France}

\date{March 17, 2016}

\begin{abstract}
We explore the computation of high-harmonic generation spectra by means of Gaussian basis sets in approaches propagating the time-dependent Schr\"odinger equation.
We investigate the efficiency of Gaussian functions specifically designed for the description of the continuum proposed by Kaufmann {\it et al.} [J. Phys. B {\bf 22}, 2223 (1989)]. 
We assess the range of applicability of this approach by studying the hydrogen atom, i.e. the simplest atom for which ``exact'' calculations on a grid can be performed.
We notably study the effect of increasing the basis set cardinal number, the number of diffuse basis functions, and the number of Gaussian pseudo-continuum basis functions 
for various laser parameters. 
Our results show that the latter significantly improve the description of the low-lying continuum states, and provide a satisfactory agreement with grid calculations 
for laser wavelengths $\lambda_0$ = 800 and 1064 nm. The Kaufmann continuum functions therefore appear as a promising way of constructing Gaussian basis sets for studying molecular electron dynamics 
in strong laser fields using time-dependent quantum-chemistry approaches.
\end{abstract}

\maketitle

\section{Introduction}

High-harmonic generation (HHG) is a highly nonlinear optical phenomenon \cite{Antoine:1996vs} of increasing interest because it can provide coherent XUV and soft X-ray radiation with attosecond
($10^{-18}$ s) durations. This property offers the opportunity to investigate unexplored research areas in atoms and molecules with unprecedented time resolution
\cite{itat04nat,goul10nat,Haessler:2010hb,cork-krau07nat,krau-ivan09rmp,Sansone:2010kb}.

The HHG optical spectrum has a distinctive shape: a rapid decrease of the intensity for the low-order harmonics consistent with perturbation theory, followed by a broad plateau region where 
the harmonic intensity remains almost constant, and then an abrupt cutoff, beyond which almost no harmonics are observed. 
The HHG process can be understood by means of semi-classical pictures, such as the celebrated three-step model~\cite{cork93prl,lewe+pra94}: (i) an electron escapes from the nuclei through tunnel ionization associated with the strong laser field, (ii) it is accelerated away by the 
laser field until the sign of the field changes, (iii) whereupon the electron is reaccelerated back to the nucleus, where it may emit a photon 
as it recombines to the ground state. A key quantity emerging from the model is the maximum energy the field can provide to the electron, $E_{\text{cutoff}} = I_\text{p} + 3.17 \, U_\text{p}$,
 where $I_\text{p}$ is the ionization potential and $U_\text{p}$ is the ponderomotive energy \cite{cork93prl,lewe+pra94}.

HHG has been studied for many years with theoretical methods solving the time-dependent Schr\"odinger equation using a real-space representation of the wave 
function~\cite{Krause+92pra,Bauer01,LeinPhysRevA.66.023805,peng-star06jcp,Han10,Bandrauk:2009ig,Gordon:2006ej,Taieb:2003db,Ruiz:2006bz}. 
These grid-based methods are taken as the numerical reference for this kind of calculations.
Indeed, these approaches have proven to be accurate enough to explain key features of atomic and molecular HHG spectra. However, grid calculations imply memory and CPU requirements that rapidly become prohibitive with increasing numbers of electrons.
Because of this limitation, multielectron systems are handled in practice via the use of effective potentials keeping a single-active electron.

By contrast, quantum-chemistry methods such as time-dependent configuration interaction (TDCI)~\cite{Kla-PRB-03,Krause:2007dm,lupp+12mol,lupp+13jcp},
 multiconfiguration time-dependent Hartree-Fock~\cite{CaiZanKitKocKreScr-PRA-05}, or time-dependent density-functional theory~\cite{ding+11jcp} using local basis functions can more easily
 handle multielectron systems such as molecules, including the treatment of electron correlation. The main problem of these methods lies in the difficulty to accurately represent 
the continuum part of the system eigenstate spectrum. Addressing this issue can be done on one-electron systems, such as the H atom, since only one electron is promoted 
into the continuum during the HHG process.

In this context, the TDCI method with a Gaussian-type orbital (GTO) basis set and
a heuristic lifetime model \cite{Klinkusch:2009iw} was recently applied to the calculation of the dipole form of the HHG spectrum for the H atom \cite{lupp+13jcp}. The role of the Rydberg and the continuum states was discussed in detail, and reasonable HHG spectra (plateau/cutoff) have been obtained, when compared with the prediction from the three-step model~\cite{cork93prl,lewe+pra94} and grid-based calculations \cite{Bandrauk:2009ig}. However, the background region, beyond the harmonic cutoff, was higher than expected and spurious harmonics were present.

A possible reason of this behavior is that the basis sets adopted in Ref.~\onlinecite{lupp+13jcp} describe Rydberg states better than the continuum ones. 
Indeed, while GTO basis sets have been successfully applied for calculations of bound-state electronic properties (even for non-linear optical properties such as second-order hyperpolarizabilities, see e.g. Ref. \onlinecite{vil10}), the inherent local nature of GTO functions makes it difficult to properly describe continuum states extending over large distances (see, e.g., Ref. \onlinecite{bof16}). In Refs.~\onlinecite{lupp+13jcp} and \onlinecite{white15}, standard GTO basis sets have been augmented with a large number of diffuse
basis functions and/or basis functions centered away from the nucleus in order to cover the large spatial extension of the time-dependent wave function. 
However, this strategy has the serious drawback of only increasing the number of Rydberg states while the number of continuum states is not substantially changed. 
This results in an unbalanced description of the Rydberg and continuum states.

Few attempts have been reported in the literature to further improve GTO basis sets for a better description of the continuum states. Kaufmann {\it et al.}~\cite{kauf+89physb} proposed 
to fit GTO basis functions to Slater-type orbital basis functions having a single fixed exponent $\zeta = 1$, supposed to be adequate for scattering calculations. 
Nestmann and Peyerimhoff~\cite{NesPey-JPB-90} proposed to fit a linear combination of GTO basis functions to a set of spherical Bessel functions, which are the spherically-adapted 
continuum eigenfunctions for zero potential. Faure {\it et al.}~\cite{faure02} extended the work of Nestmann and Peyerimhoff to the possibility of fitting a linear combination of GTO basis 
functions to a set of Coulomb continuum functions (i.e., the continuum eigenfunctions obtained in presence of the Coulomb potential $-Z/\vert {\bf r}\vert$, with $Z$ the nuclear charge). 
Finally, some hybrid methods have also been proposed, combining Gaussian functions with finite-element/discrete-variable representation techniques~\cite{Yip2014pra}
or with B-spline basis sets~\cite{mara14pra}. Note that an alternative approach to Gaussian basis sets is given by the use of Sturmian functions~\cite{frap10,gran14}.\\

In this article, we study the merits of the Gaussian continuum basis functions proposed by Kaufmann \emph{et al.} \cite{kauf+89physb} for calculating the HHG spectra in atomic hydrogen within
the TDCI framework. While the present results are focused on HHG, our work is relevant for the calculation of any property involving electronic transitions toward the continuum such as, e.g.,
photoionization cross sections \cite{Reinhardt:1979} or above-threshold ionization rates \cite{Klinkusch:2009iw}.
The paper is organized as follows. We first describe the theory and give computational details. 
We then present and discuss our results. In particular, we show velocity HHG spectra extracted from the dipole, velocity, and acceleration power spectra calculated for different laser intensities, 
and basis sets. We study in detail the effect of increasing the basis set cardinal number, the number of diffuse basis functions, and the number of Gaussian continuum basis 
functions. We directly compare our results with data from grid calculations, for three values of the laser intensity and two values of the laser wavelength, and adjust the heuristic lifetime model.
 Finally, we conclude with final comments and perspectives. 
Unless otherwise noted, Hartree atomic units, i.e. $\hbar=m_\text{e}=e^2/(4\pi\epsilon_0)=1$, are used throughout the paper.

\section{Theoretical method}
\label{theory}

The time-dependent Schr\"odinger equation for the H atom in an external time-dependent uniform electric field ${\bf E}(t)$ in the length gauge is 
\begin{equation}
i\frac{\partial \vert \Psi(t) \ket }{\partial t} = \left( \hat{H}_0 + \hat{V}(t) \right) \vert \Psi(t) \ket,
\label{tdse}
\end{equation}
where $H_0({\bf r}) = -\nabla^2/2 -1/\vert {\bf r}\vert$ is the time-independent field-free Hamiltonian and $V({\bf r},t) = {\bf r}  \cdot {\bf E}(t)$ is the interaction potential between the atom and the field in the semiclassical dipole approximation. 
We consider the case of an electric field ${\bf E}(t)$ linearly polarized along the $z$-axis, representing a laser pulse, 
\begin{equation}
{\bf E}(t) = E_0 {\bf n}_z \sin(\omega_{0} t + \phi) f(t),
\label{e-field}
\end{equation}
where $E_0$ is the maximum field strength, ${\bf n}_z$ is the unit vector along the $z$ axis, $\omega_{0}$ is the carrier frequency, $\phi$ is the carrier-envelope phase, and $f(t)$ is the envelope function chosen as
\begin{equation} 
f(t) = 
\begin{cases}  
\cos^{2}( \frac{\pi}{2\sigma} (t - \sigma ) )   & \text{if }  0 \le t \le 2\sigma,\\ 
0 & \text{otherwise},
\end{cases} 
\label{enve}
\end{equation}
where $\sigma$ is the full width at half maximum of the field envelope.

The target quantity to be computed is the power spectrum $P_{\xi}(\omega)$ defined as
\begin{equation}
P_{\xi}(\omega) = \bigg|  \frac{1}{t_\f - t_\i} \int^{t_\f}_{t_\i} \bra \Psi(t) \vert \hat{\xi} \vert \Psi(t) \ket e^{-i \omega t} \d t \bigg|^{2},
\label{FFT}
\end{equation}
where $t_\i$ and $t_\f$ are the initial and final propagation times. In Eq.~(\ref{FFT}), the operator $\hat{\xi}$ can be either equal to the position operator $\hat{z}$, or to the velocity operator $\hat{v}_z=- i [\hat{z},\hat{H}(t)]$, or to the acceleration operator $\hat{a}_z = -i [\hat{v}_z,\hat{H}(t)]$ (where $\hat{H}(t) = \hat{H}_0 + \hat{V}(t)$ is the total time-dependent Hamiltonian), defining three different forms of the power spectrum: the dipole $P_z(\omega)$,
the velocity $P_{v_{z}}(\omega)$, and the acceleration $P_{a_{z}}(\omega)$ forms. According to recent works~\cite{Die-PRA-08,BagMad-JPB-11}, the velocity form $P_{v_{z}}(\omega)$ best 
represents the HHG spectrum of a single atom or molecule. The three forms are related to each other by (see Appendix~\ref{app:powerspectrum}):
\begin{equation}
\omega^2 P_{z}(\omega) \approx P_{v_z}(\omega) \approx \frac{1}{\omega^2} P_{a_z}(\omega).
\label{eq:equi}
\end{equation}
In this work, we always show the same quantity, i.e. the velocity HHG spectrum, either extracted directly from the velocity power spectrum, or indirectly from the dipole or 
the acceleration power spectrum with the appropriate frequency factors following Eq.~(\ref{eq:equi}).

\subsection{Time-propagation scheme}
    
The time-dependent Schr\"odinger equation is solved using the TDCI method (see, e.g., Refs.~\onlinecite{Kla-PRB-03,Krause:2007dm,lupp+12mol,lupp+13jcp}) applied to the special case of the H atom.
The wave function $ \vert \Psi(t) \ket $ is expanded in the discrete basis of the eigenstates $\{\vert \psi_{k}\ket\}$ of the field-free Hamiltonian $\hat{H}_0$ (projected in the same basis),
composed of the ground state ($k=0$) and all the excited states ($k>0$)
\begin{equation}
\vert \Psi(t) \ket= \sum_{k \geq 0} {c}_{k}(t) \vert \psi_{k}\ket,
\label{tdcis_state}
\end{equation} 
where ${c}_{k}(t)$ are time-dependent coefficients. Inserting Eq.~(\ref{tdcis_state}) into Eq.~(\ref{tdse}), and projecting on the eigenstates $\bra \psi_{l}\vert$, gives the evolution equation 
\begin{equation}
i\frac{\d {\bf c}(t)}{\d t} = \left( {\bf H}_0 + {\bf V}(t) \right) {\bf c}(t),
\label{CIequation}
\end{equation}
where ${\bf c}(t)$ is the column matrix of the coefficients ${c}_{k}(t)$, ${\bf H}_0$ is the diagonal matrix of elements ${\bf H}_{0,lk} = \bra \psi_l |\hat{H}_0 | \psi_k \ket = E_k \delta_{lk}$ 
(where $E_k$ is the energy of the eigenstate $k$), and ${\bf V}(t)$ is the non-diagonal matrix of elements ${\bf V}_{lk}(t)= \bra \psi_l |\hat{V}(t) | \psi_k \ket$. The initial wave function 
at $t=t_\i=0$ is chosen to be the field-free ground state, i.e. $c_k(t_\i) = \delta_{k0}$. To solve Eq.~(\ref{CIequation}), time is discretized and the simple split-propagator approximation is used to separate the contributions of the field-free Hamiltonian ${\bf H}_0$ and the atom-field interaction ${\bf V}(t)$
\begin{equation}
{\bf c}(t + \Delta t) \approx  e^{-i {\bf V}(t) \Delta t}  e^{-i {\bf H}_0 \Delta t}  {\bf c}(t),
\label{coefficients}
\end{equation}
where $\Delta t$ is a small time step. Since the matrix ${\bf H}_0$ is diagonal, $e^{-i {\bf H}_0 \Delta t}$ is a diagonal matrix of elements $e^{-i E_k \Delta t}\delta_{lk}$. The exponential of the non-diagonal matrix ${\bf V}(t)$ is calculated as 
\begin{equation}
e^{-i {\bf V}(t) \Delta t} = {\bf U}^{\dagger} \; e^{-i {\bf V}_\text{d} (t) \Delta t} \; {\bf U},
\end{equation}
where ${\bf U}$ is the unitary matrix describing the change of basis between the original eigenstates of $\hat{H}_0$ and a basis in which the atom-field interaction $\hat{V}(t)$ is diagonal, i.e. ${\bf V}(t) = {\bf U}^{\dagger} {\bf V}_\text{d} (t) { \bf U} = {\bf E}(t) \cdot {\bf U}^{\dagger} {\bf r}_\text{d} { \bf U} $ where ${\bf V}_\text{d} (t) = {\bf E}(t) \cdot {\bf r}_\text{d}$ is the diagonal atom-field interaction matrix and ${\bf r}_\text{d}$ is the diagonal representation matrix of the position operator. Since the time dependence is simply factorized in a multiplicative function independent of ${\bf r}$, the unitary matrix ${\bf U}$ is time-independent and can be calculated once and for all before the propagation.

Once the time-dependent coefficients are known, it is possible to calculate the time-dependent dipole, velocity, or acceleration as
\begin{equation}
\xi(t) = \bra \Psi(t) \vert \hat{\xi} \vert \Psi(t) \ket = \sum_{l,k} c_l^*(t) c_k(t) \bra \psi_l \vert \hat{\xi} \vert \psi_k \ket,
\end{equation}
which, after taking the square of its Fourier transform, leads to the corresponding power spectrum of Eq.~(\ref{FFT}).

\subsection{Gaussian basis sets}

The field-free states (simply corresponding to the atomic orbitals for the H atom) are expanded on a Gaussian basis set,
\begin{equation}
\vert \psi_{k}\ket = \sum_\mu d_{\mu,k} \vert \chi_\mu \ket,
\end{equation} 
where $\{\chi_\mu\}$ are real-valued GTO basis functions centered on the nucleus. In spherical coordinates ${\bf r} = (r,\theta,\phi)$,
\begin{equation}
\bra {\bf r} \vert \chi_\mu \ket = N_{\alpha_{\mu},\ell_\mu} r^{\ell_\mu} e^{-\alpha_{\mu} r^2} S_{\ell_\mu,m_\mu}(\theta,\phi),
\end{equation} 
where $N_{\alpha_{\mu},\ell_\mu}$ is a normalization constant, $\alpha_{\mu}$ are exponents, $S_{\ell,m}(\theta,\phi)$ are real spherical harmonics.

We built the Gaussian basis set starting from the Dunning basis sets~\cite{Dun-JCP-89}, adding first diffuse GTO functions to describe the Rydberg states, and a special set of GTO functions adjusted to represent low-lying continuum states. For the latter, we follow Kaufmann {\it et al.}~\cite{kauf+89physb} who proposed to fit GTO basis functions to Slater-type orbital basis functions having a single fixed exponent $\zeta = 1$. For each angular momentum $\ell$, Kaufmann {\it et al.} found a sequence of optimized GTO exponents which are well represented by the simple formula~\cite{kauf+89physb}
\begin{equation}
\alpha_{\ell,n} = \frac {1}  {4(a_\ell \; n + b_\ell)^2},
\label{kauf_sca}
\end{equation}
where $n=1,2,3,...$ is not associated to the quantum principal number but is just an index identifying a given value in the list of all exponents for a fixed $\ell$, and the parameters $a_\ell$ and $b_\ell$ are given in Table 2 of Ref.~\onlinecite{kauf+89physb}. 
The GTO basis functions obtained with these exponents (collected in Table \ref{tab:continuum})
will be in the following referred to as ``Gaussian continuum functions'' or ``Kaufmann (K) functions''.

\begin{table}
\caption{Exponents $\alpha_{\ell,n}$ [see Eq. (\ref{kauf_sca})] of the Gaussian functions for describing the continuum proposed by Kaufmann {\it et al.} \cite{kauf+89physb} and used in the present work for $n=1,...,8$ and $\ell=0,1,2$.
\label{tab:continuum}}
\begin{tabular}{c   c   c   c }
\hline\hline
$n$ & $\ell=0$ & $\ell=1$ & $\ell=2$  \\
\hline 
1 &  0.245645       &  0.430082     & 0.622557  \\
2 &  0.098496       &  0.169341     & 0.242160   \\
3 &  0.052725       &  0.089894     & 0.127840   \\
4 &  0.032775       &  0.055611     & 0.078835   \\
5 &  0.022327       &  0.037766     & 0.053428 \\
6 &  0.016182       &  0.027312     & 0.038583  \\
7 &  0.012264       &  0.020666     & 0.029163 \\
8 &  0.009615       &  0.016181     & 0.022815 \\
\hline\hline
\end{tabular}
\end{table}

\subsection{Finite lifetime model}

The GTO basis set incompleteness is responsible for an incorrect description of the continuum eigenfunctions. They decay too fast for large $r$, which prevents the description of 
the above-threshold ionization and leads to unphysical reflections of the wave function in the laser-driven dynamics. To compensate for this, we use the heuristic lifetime model of
 Klinkusch {\it et al.}~\cite{Klinkusch:2009iw} which consists in interpreting the approximate field-free eigenstates $\psi_k$ above the ionization threshold (taken as the zero energy reference) 
as non-stationary states 
and thus replacing, in the time propagation, the energies $E_k$ by complex energies $E_k - i \Gamma_k/2$, where $\Gamma_k$ is the inverse lifetime of state $k$. For the special case of the H atom 
the $\Gamma_k$ are chosen as ~\cite{Klinkusch:2009iw}
\begin{equation}
\label{heuri}
\Gamma_k  = \left\{
\begin{array}{rl}
0 & \text{if}\quad E _k < 0,\\
\sqrt{2 E_k}/d  & \text{if}\quad  E_k >0,
\end{array} \right.
\end{equation}
where $d$ is an empirical parameter representing the characteristic escape length that the electron in the state $k$ is allowed to travel during the lifetime $1/\Gamma_k$. 
These complex energies are used in the propagation described by Eq.~(\ref{coefficients}), in the field-free Hamiltonian matrix ${\bf H}_0$. The heuristic lifetime model is a 
simple alternative to using complex scaling~\cite{TelSosRozChu-PRA-13,book:Moiseiev}, a complex-absorbing potential~\cite{Greenman:2010je,Krause14},
 or a wave-function absorber~\cite{Krause+92pra}.

In this work, we also introduce and test a modified version of the original heuristic lifetime model. In this version, two different values of the escape length,
$d_0$ and $d_1$, are used to increase the flexibility in the definition of the finite lifetimes, adapted to the present context of HHG. A large value of $d_0$ (small value of $\Gamma_k$) is used for 
all the above-ionization-threshold states with positive energy below the energy cutoff of the three-step model $E_\text{cutoff}$, while a smaller $d_1$ (larger $\Gamma_k$) is used for the continuum 
states with energies above $E_\text{cutoff}$, which are not expected to contribute to HHG. This allows us to better retain the contribution of low-energy continuum states
 for the recombination step of the HHG process.

\section{Computational details}
\label{computational}

\begin{table}
\caption{Physical parameters relevant to HHG for the H atom with two laser wavelengths $\lambda_0= 800$ and $1064$ nm and
three laser intensities $I= 5 \times$10$^{13}$ W/cm$^{2}$, $10^{14}$ W/cm$^{2}$, and $2\times$10$^{14}$ W/cm$^{2}$: Keldysh parameter $\gamma=\sqrt{I_\text{p}/(2 U_\text{p})}$ \cite{kel67},
 ponderomotive energy $U_{\text{p}} = E_0^{2} / (4\omega_{0}^{2})$ (in hartree), energy cutoff in the three-step model $E_\text{cutoff}= I_\text{p} + 3.17 U_{\text{p}}$ (in hartree) 
where $I_\text{p}=0.5$ hartree is the ionization potential, 
harmonic cutoff in the three-step model $N_{\text{cutoff}} = E_\text{cutoff}/\omega_{0}$, and
maximum electron excursion distance in the continuum $R_{\text{max}} = 2E_0 / \omega_0^2$ (in bohr) in the three-step model.
\label{tab:lasers}}
\begin{tabular}{c  c  c  c }
\hline \hline 
$I$ & 5$\times$10$^{13}$ W/cm$^2$ & 10$^{14}$ W/cm$^2$ & 2$\times$10$^{14}$ W/cm$^2$ \\
\hline  
$\lambda_0=800$ nm\\
$\gamma$                &  1.51  & 1.06      & 0.76 \\
$U_{\text{p}}$          &  0.11  & 0.22      & 0.44 \\ 
$E_\text{cutoff}$       &  0.85  & 1.20      & 1.89 \\
$N_{\text{cutoff}}$  &  15 &  21 & 33 \\
$R_{\text{max}}$     & 23 & 33 & 46 \\
\\
$\lambda_0=1064$ nm\\
$\gamma$                &  1.13  &   0.79    & 0.57 \\
$U_{\text{p}}$          &  0.19  &   0.40    & 0.78 \\ 
$E_\text{cutoff}$       &  1.10  &   1.77    & 2.97  \\
$N_{\text{cutoff}}$  &  26 &  41 &  69 \\
$R_{\text{max}}$     & 41 & 59 & 82 \\
\hline \hline 
\end{tabular}
\end{table}

The field-free calculations are performed using a development version of the \textsc{Molpro} software package~\cite{MOLPRO_brief} from which all the electronic energies, as well as the dipole, velocity, and
acceleration matrix elements over the electronic states have been obtained. 
The external code \textsc{light} \cite{lupp+13jcp} is used to perform the time-propagation using a time step $\Delta t$ = 2.42 as (0.1 a.u.) and the Fourier transformations with a Hann window 
function. An escape length $d$=1.41 bohr is used for the original heuristic lifetime model, while $d_0=50$ bohr and $d_1=0.1$ bohr are chosen for the modified version 
of the heuristic lifetime model as explained in the section Results and Discussion.

Correlation-consistent $N$-aug-cc-pV$X$Z~\cite{lupp+13jcp} basis sets are used, where $X$ is the cardinal number ($X=$ T, Q, 5) connected to the maximum angular
momentum ($L_\text{max}=X-1$ for the H atom), and $N$ is the number of shells of diffuse functions for each angular momentum. We only employ $N=6$ or $N=9$ because $N=6$ can be
considered as the minimum augmentation needed to reasonably describe HHG spectra for the H atom~\cite{lupp+13jcp}. In particular, the 6-aug-cc-pVTZ basis set describes up to $(n=3)$-shell 
Rydberg states, 6-aug-cc-pVQZ up to $(n=4)$-shell Rydberg states, and 6-aug-cc-pV5Z up to $(n=5)$-shell Rydberg states. Furthermore, we investigate the effect of adding to the 6-aug-cc-pVTZ
basis set 3, 5, and 8 Gaussian continuum functions (or K functions) for each angular momentum. The extra diffuse and continuum Gaussian functions 
are uncontracted.

For comparison, we also perform accurate grid calculations in the length gauge. The wave function is expanded on a set of spherical harmonics $Y_{\ell,m}(\theta,\phi)$ up to $\ell = 128$, and the resulting coupled equations are discretized
on a radial grid with a step size of $\Delta r = 0.25$ bohr (see Ref. \onlinecite{Krause+92pra}). 
A box size of 256 bohr is used with a mask function~\cite{Krause+92pra,Taieb:2003db} at 200 bohr to absorb the part of the wave function accounting for ionized electrons that will not rescatter 
towards the nucleus. The mask function multiplying the wave function at each time step has been chosen to be cos($r$)$^{1/8}$, which  is effective in modeling the ionization
\cite{Krause+92pra}. 
The time step used is $\Delta t$=0.65 as (0.027 a.u.). The grid-based calculations, being converged with respect to the parameters mentioned above, 
represent the numerical reference for the current GTO results. We note that performing the grid-based calculations takes hours on a standard workstation,
while the field-free and time-propagation calculations in the GTO basis sets take only a few minutes.

Unless otherwise noted, the calculations are done with the carrier laser frequency $\omega_0$ = 1.550 eV ($\lambda_0 = 800$ nm), corresponding to a Ti:sapphire laser.
 For the comparison with the grid calculations, we also use the laser frequency $\omega_0$ = 1.165 eV ($\lambda_0 = 1064$ nm) for which higher-energy regions are probed. 
The pulse duration is $2\sigma=20$ oc where 1 optical cycle (oc) is $2\pi/\omega_{0}$ (110.23 a.u.).
 We use three peak laser intensities $I= (\varepsilon_0 c/2) E_0^2$: $I = 5\times10^{13}$  W/cm$^2$, $I=10^{14}$  W/cm$^2$, and $I= 2\times10^{14}$ W/cm$^2$. 
We have thus chosen a range of intensities encompassing the over-barrier ionization threshold (i.e. the critical intensity above which the electron can classically overstep the barrier)
of hydrogen, $I_\text{b}=1.4 \times 10^{14}$ W/cm$^{2}$. We can therefore study the performance of our method in realistic conditions for which HHG progressively becomes less pronounced
with increasing laser intensity. The physical parameters relevant to HHG are reported in Table~\ref{tab:lasers}.

\section{Results and Discussion} 
\label{resultsdiscussion}
  
We start by studying the performance of several Gaussian basis sets for the calculation of HHG spectra of the H atom, continuing the previous work of Luppi and Head-Gordon~\cite{lupp+13jcp}.
The optimal basis set including Gaussian continuum functions is then used for a direct comparison with reference HHG spectra from grid calculations.

\subsection{Time-dependent dipole, velocity, and acceleration}    
 
\begin{figure}
\begin{center}
\includegraphics[width=3in,angle=0]{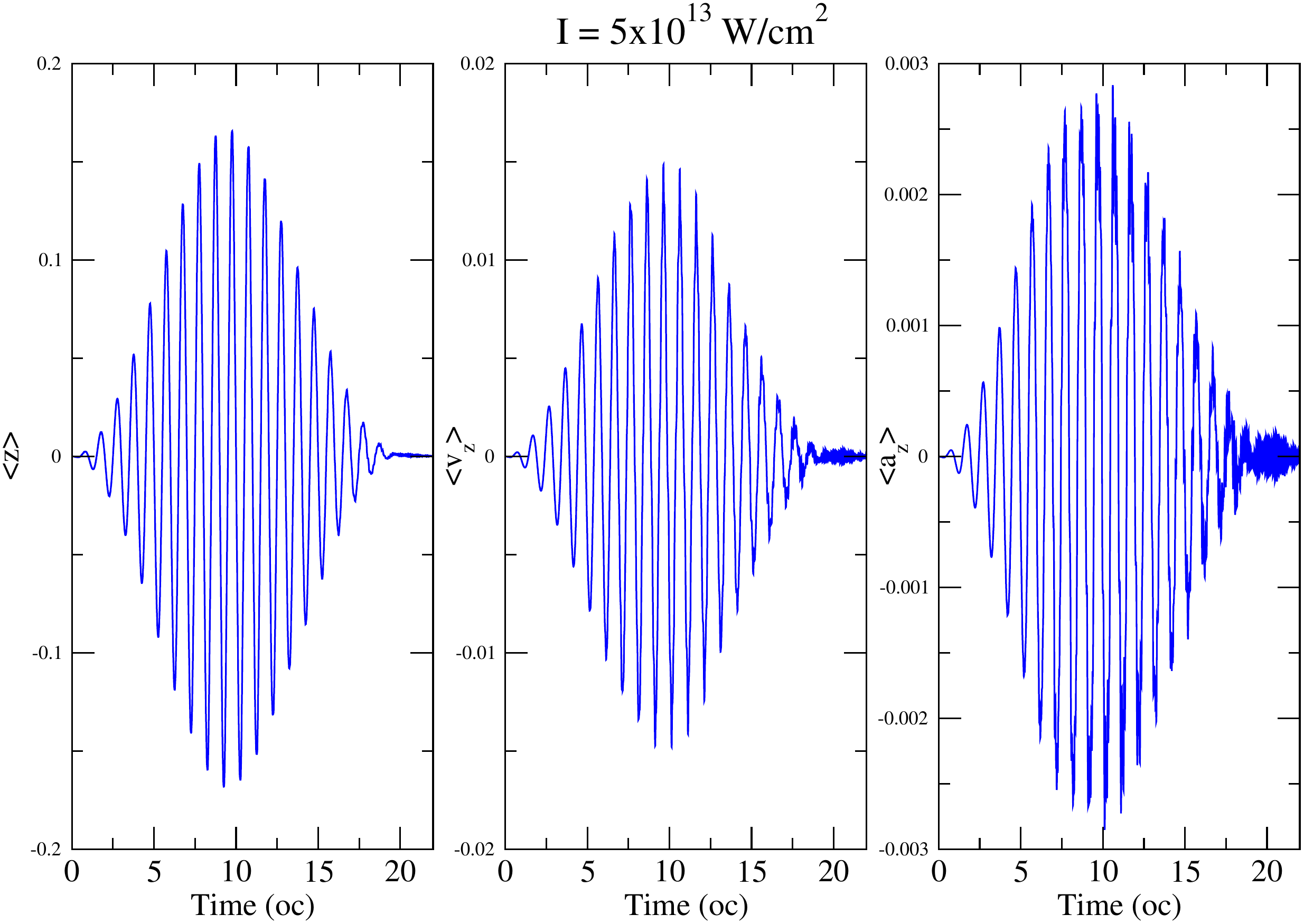} \\
\includegraphics[width=3in,angle=0]{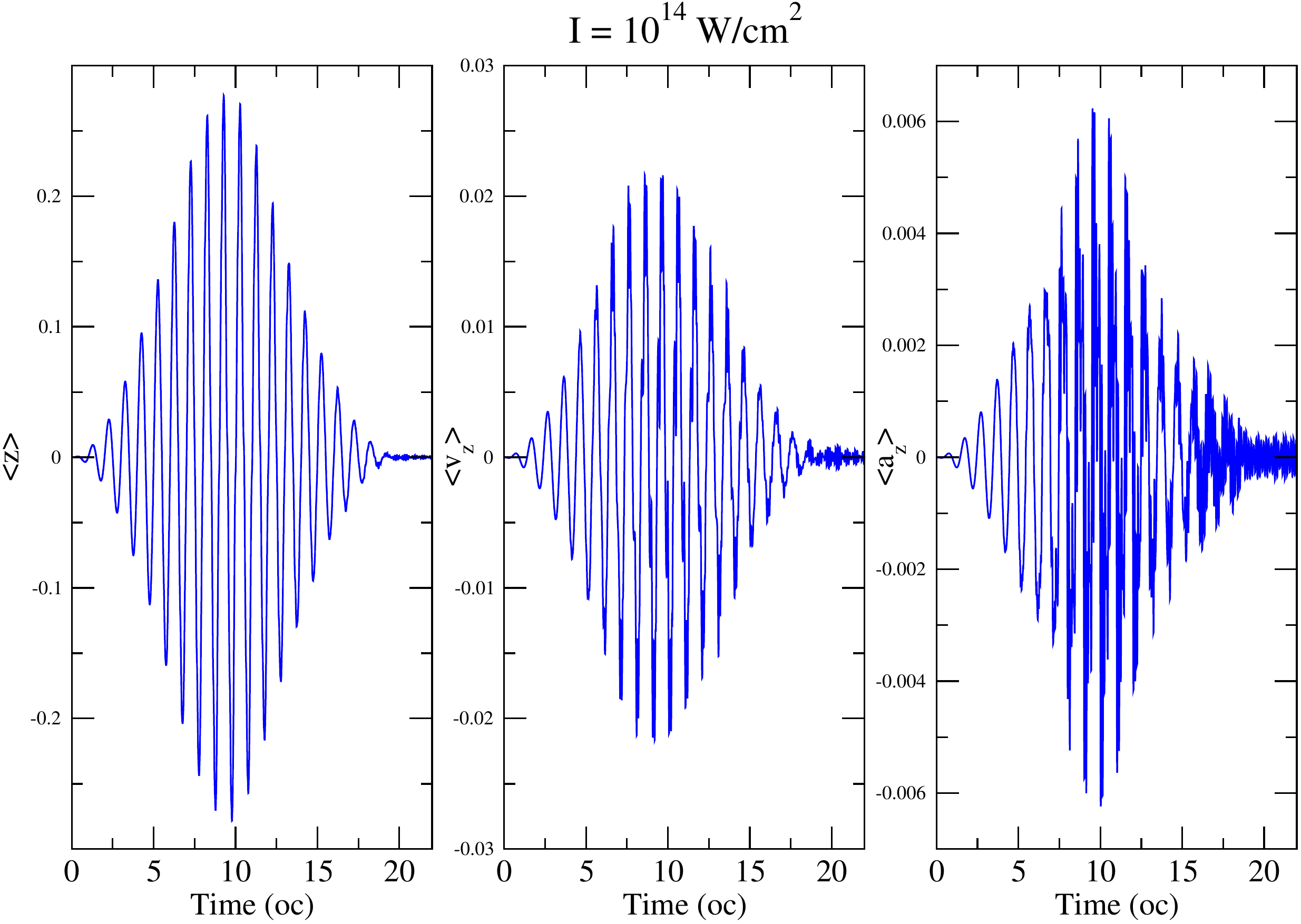} \\
\includegraphics[width=3in,angle=0]{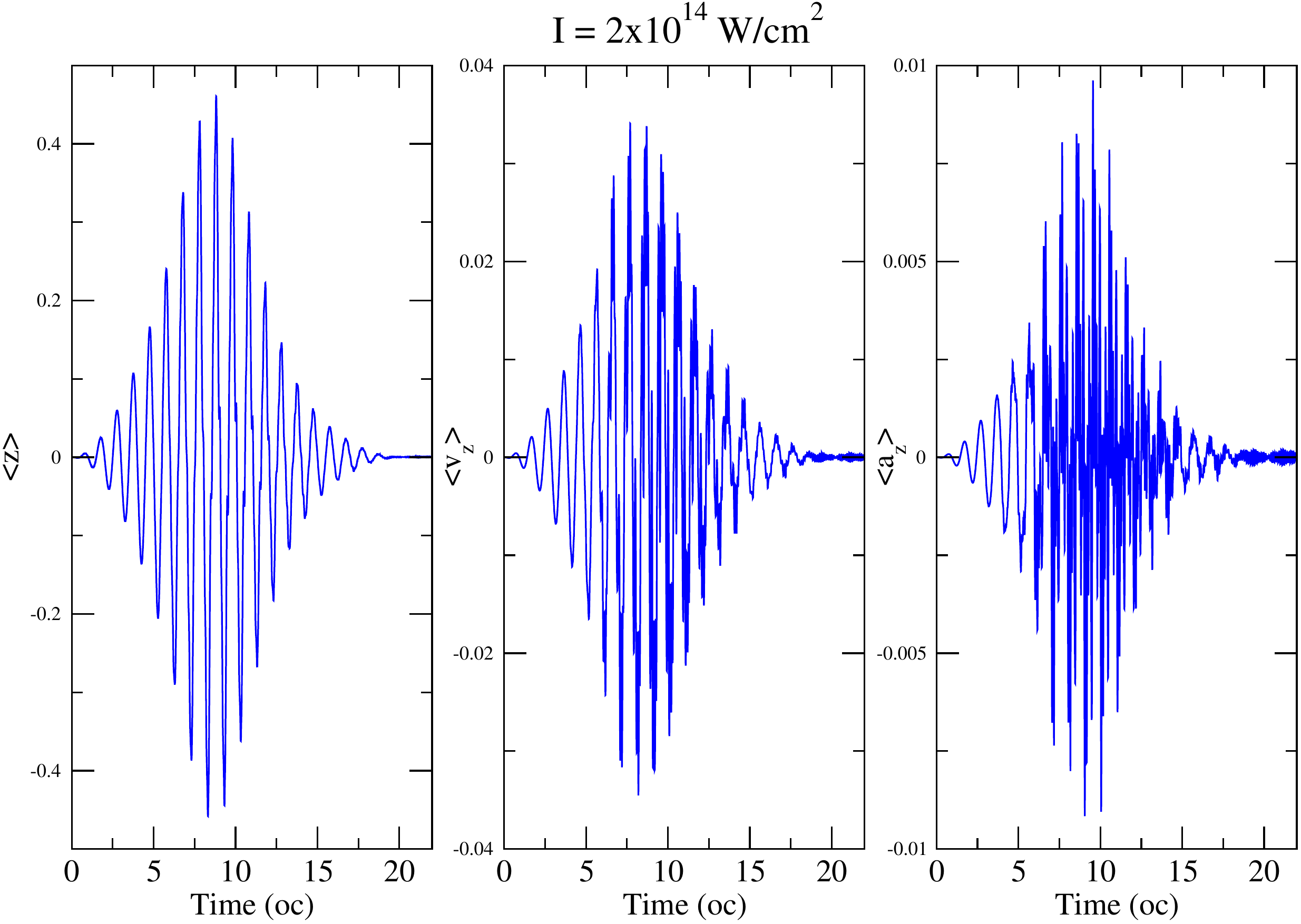}
\caption{Time-dependent dipole $z(t)$ (left), velocity $v_z(t)$ (middle), and acceleration $a_z(t)$ (right) calculated with the 6-aug-cc-pVTZ basis set for laser
 intensities $I = 5 \times$10$^{13}$ W/cm$^{2}$ (top), $I=10^{14}$ W/cm$^{2}$ (middle), and $2\times$10$^{14}$ W/cm$^{2}$ (bottom). \label{fig1}} 
\end{center}
\end{figure}  

We have reported on Figure~\ref{fig1} the time evolution of the dipole $z(t)$, the velocity $v_z(t)$, and the acceleration $a_z(t)$ with the 6-aug-cc-pVTZ basis set for the three 
laser intensities. The evolution of $z(t)$, $v_z(t)$, and $a_z(t)$ follows the shape of the laser field given in Eq.~(\ref{enve}), with the shape of their envelopes changing with the intensity
of the pulse. Note that $v_z(t)$ is one order of magnitude smaller than $z(t)$ and its oscillations have a finer structure. Similarly, $a_z(t)$ is one order of magnitude smaller 
than $v_z(t)$ and has even more structured oscillations. Even though some fast oscillations are still present after the laser is switched off due to the population of electronic excited states,
the conditions $z(t_\text{f})=0$ and $v_{z}(t_\text{f})=0$ (see Appendix~\ref{app:powerspectrum}) are approximately fulfilled, which will allow us to use Eq.~(\ref{eq:equi}). Our results are in reasonable agreement with the results of 
Bandrauk {\it et al.}~\cite{Bandrauk:2009ig} and those of Han and Madsen~\cite{Han10} who used grid-based methods. Similar findings have been reported for the He atom in a low-field regime 
using time-dependent Hartree-Fock and time-dependent Kohn-Sham with Gaussian basis sets~\cite{ding+11jcp}.

\subsection{Dipole, velocity, and acceleration forms of the HHG spectrum}

\begin{figure}
\begin{center}
\includegraphics[width=3.5in,angle=0]{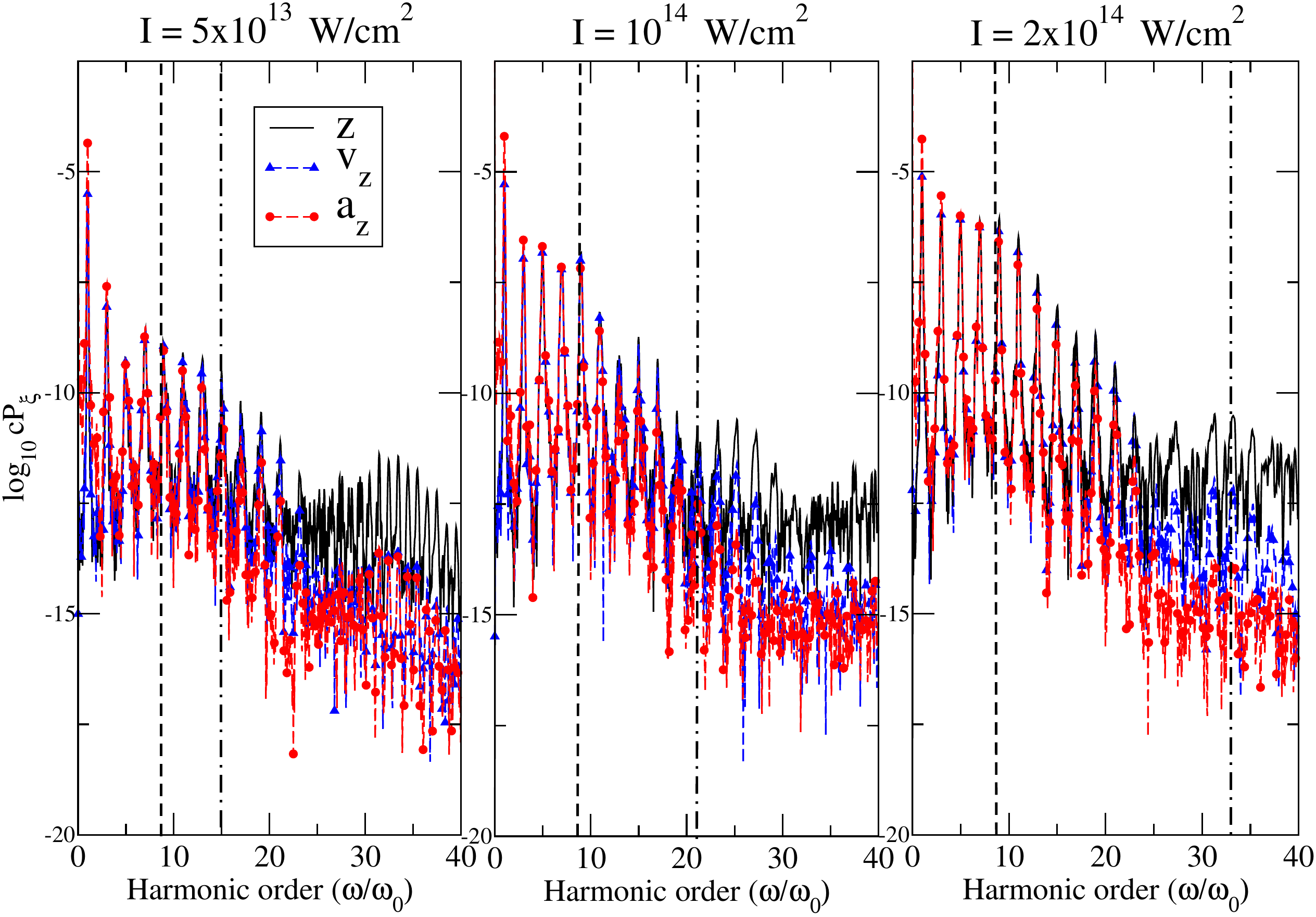}
\caption{Velocity HHG spectra of the H atom extracted from the dipole power spectrum $\omega^2 P_z(\omega)$ (i.e. $c=\omega^2$), the velocity power spectrum $P_{v_z}(\omega)$ (i.e. $c=1$), and the acceleration power spectrum $P_{a_z}(\omega)/\omega^2$ (i.e. $c=1/\omega^2$)
calculated with the 6-aug-cc-pVTZ basis set and laser intensities $I = 5 \times$10$^{13}$ W/cm$^{2}$, $I=10^{14}$ W/cm$^{2}$, and $I = 2\times$10$^{14}$ W/cm$^{2}$.
The ionization threshold ($I_\text{p}/\omega_0$, vertical dashed line) and the harmonic cutoff in the three-step model $N_{\text{cutoff}}$ (vertical dot-dashed line) are also shown.
\label{fig2}} 
\end{center}
\end{figure} 

In Figure~\ref{fig2} the velocity HHG spectrum, extracted from the dipole, velocity, and acceleration power spectra according to Eq.~(\ref{eq:equi}), calculated with the 6-aug-cc-pVTZ basis set 
and the three laser intensities are shown. 
The typical form of the HHG spectrum (plateau/cutoff/background) is obtained. We note that the harmonic peaks that we obtained are sharper than those calculated by Bandrauk \emph{et al.}~\cite{Bandrauk:2009ig} 
based on a direct propagation of the time-dependent Schr\"odinger equation on a grid. 

The dipole, velocity, and acceleration formulations of the velocity HHG spectrum give similar spectra in the plateau region, but different backgrounds beyond the harmonic cutoff. In particular, the HHG spectrum
calculated from the dipole power spectrum
presents a higher background than the HHG spectra calculated from the velocity and acceleration power spectra, in agreement with the calculations of Bandrauk \emph{et al.}~\cite{Bandrauk:2009ig}. These differences reflect the
sensitivity to the basis set. Indeed, the expectation value of the dipole operator probes the time-dependent wave function in spatial regions further away from the nucleus than the expectation values
of the velocity and acceleration operators do. In the following, since the dipole is the most difficult to converge with our basis set
 we will focus on the basis set convergence of the (velocity) HHG spectrum computed from the dipole power spectrum.

\subsection{Effect of the cardinal number of the basis set and the number of diffuse basis functions} 
\label{ang}

We first analyze the effect of the basis-set cardinal number $X$, before examining the effect of adding Gaussian continuum basis functions in Sec.~\ref{conti}. 
We use the following series of basis sets: 6-aug-cc-pVTZ (s, p, and d shells), 6-aug-cc-pVQZ (s, p, d, and f shells), and 6-aug-cc-pV5Z (s, p, d, f, and g shells). The number of total, bound
(i.e., energy below 0), and continuum (i.e., energy above 0) states, and the maximum energy obtained with these basis sets are reported in the upper half of Table~\ref{angmom}. 
Going from 6-aug-cc-pVTZ to 6-aug-cc-pV5Z the total number of states increases considerably, from 68 to 205. The percentage of continuum states also tends to increase with the cardinal number. However, these added continuum states are not necessarily in the energy range relevant to the HHG spectrum. Indeed, the maximum energies obtained are 3.45 hartree for 6-aug-cc-pVTZ, 7.74 hartree for 6-aug-cc-pVQZ, and 15.94 hartree for 6-aug-cc-pV5Z, while the maximal kinetic energy that can be transmitted to the electron ($E_\text{cutoff}-I_\text{p}$) in the three-step model are between 0.35 and 2.47 hartree for the parameters considered (see Table~\ref{tab:lasers}).

\begin{table}
\caption{Number of total, bound, and continuum states and the maximum energy $E_\text{max}$ (in hartree) obtained with the 6-aug-cc-pVTZ, 6-aug-cc-pVQZ, and 6-aug-cc-pV5Z basis sets, as well as
 with the 6-aug-cc-pVTZ+3K, 6-aug-cc-pVTZ+5K, and 6-aug-cc-pVTZ+8K basis sets. The percentages of bound and continuum states are indicated in parenthesis.\label{angmom}}
\begin{tabular}{l>{\raggedleft\let\newline\\\arraybackslash}m{10mm}
                 >{\raggedleft\let\newline\\\arraybackslash}m{05mm}
                 >{\raggedleft\let\newline\\\arraybackslash}m{05mm}
                 >{\raggedleft\let\newline\\\arraybackslash}m{05mm}
                 >{\raggedleft\let\newline\\\arraybackslash}m{05mm}
                 >{\raggedleft\let\newline\\\arraybackslash}m{10mm}}
\hline\hline 
                 & Total & \multicolumn{2}{>{\centering\let\newline\\\arraybackslash}m{16mm}}{Bound}
                         & \multicolumn{2}{>{\centering\let\newline\\\arraybackslash}m{16mm}}{Continuum}  & $E_\text{max}$\\
\hline 
6-aug-cc-pVTZ    &    68 & 42 & (62\%) &  26 & (38\%) &  3.45 \\
6-aug-cc-pVQZ    &   126 & 63 & (50\%) &  63 & (50\%) &  7.74 \\
6-aug-cc-pV5Z    &   205 & 90 & (44\%) & 115 & (56\%) & 15.94 \\ 
\hline
6-aug-cc-pVTZ+3K &    95 & 42 & (44\%) &  53 & (56\%) &  6.31 \\
6-aug-cc-pVTZ+5K &   113 & 46 & (41\%) &  67 & (59\%) &  6.68 \\
6-aug-cc-pVTZ+8K &   140 & 51 & (36\%) &  89 & (64\%) &  6.93 \\
\hline\hline 
\end{tabular}
\end{table}

\begin{figure}
\begin{center}
\includegraphics[width=3.5in,angle=0]{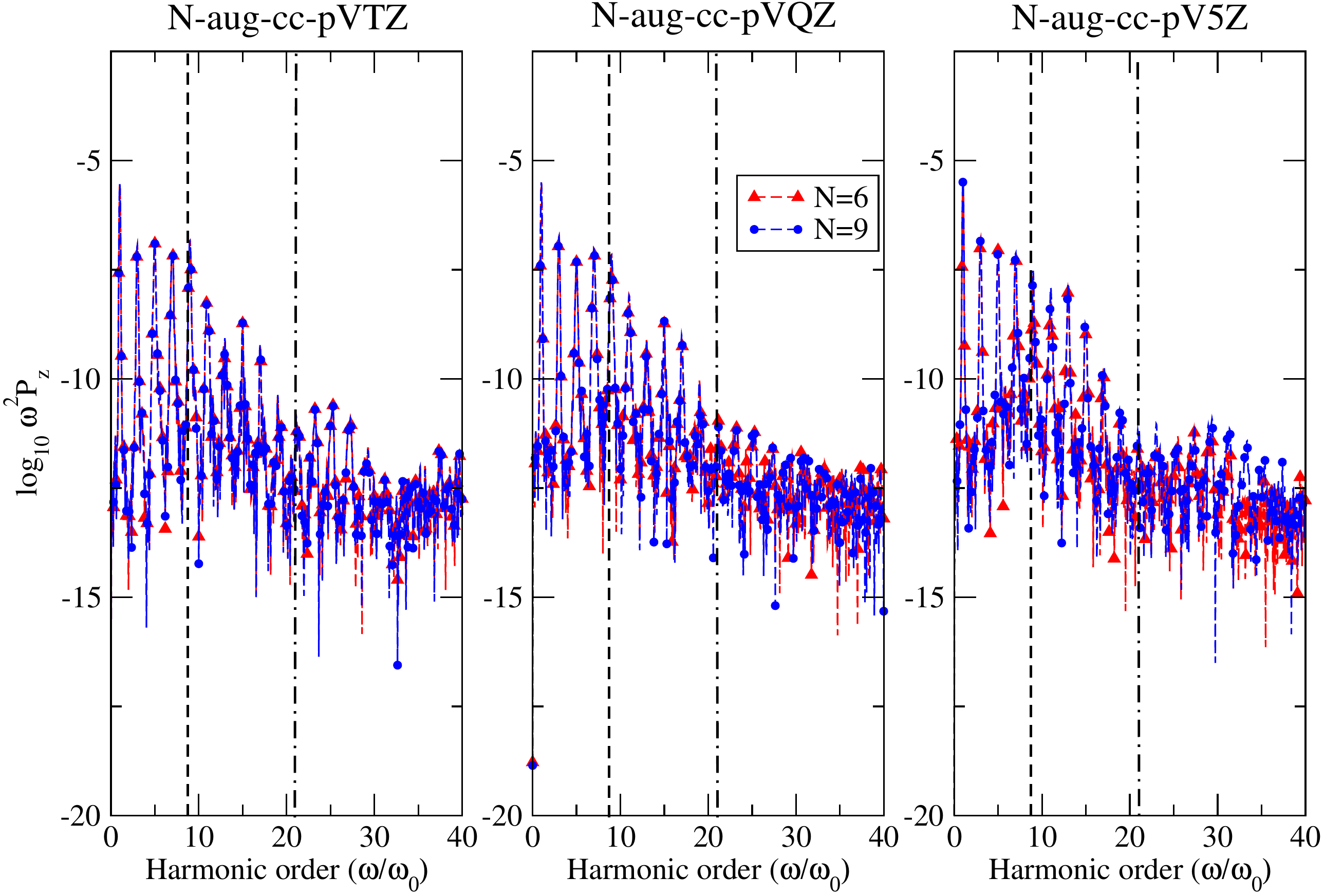} \\
\caption{Velocity HHG spectrum of the H atom extracted from the dipole power spectrum $\omega^2 P_z(\omega)$ calculated with the 6-aug-cc-pV$X$Z and 9-aug-cc-pV$X$Z basis sets with $X$= T (left), Q (middle) and 5 (right). 
The laser intensity is $I=10^{14}$ W/cm$^{2}$.
The ionization threshold ($I_\text{p}/\omega_0$, vertical dashed line) and the harmonic cutoff in the three-step model $N_{\text{cutoff}}$ (vertical dot-dashed line) are also shown.
\label{fig3}} 
\end{center}
\end{figure}

In Figure~\ref{fig3}, we compare the velocity HHG spectrum extracted from the dipole power spectrum for the 6-aug-cc-pVTZ, 6-aug-cc-pVQZ, and 6-aug-cc-pV5Z basis sets for the laser intensity $I=10^{14}$ W/cm$^{2}$. 
The three basis sets give very similar results, in the plateau as well as beyond the harmonic cutoff. We thus conclude that the HHG spectrum is not strongly affected by the cardinal number $X$ of the basis set and therefore, in the following, 
we will use a triple-zeta ($X=$ T) basis set. In Figure~\ref{fig3}, we also compare the spectra calculated using the $N$-aug-cc-pV$X$Z basis sets with $N=6$ and $N=9$. The results show that the
convergence in terms of diffuse basis functions is achieved with 6 diffuse shells.

\subsection{Effect of the Gaussian continuum basis functions} 
\label{conti}

The sensitivity of the HHG spectrum to the cardinal number and to the number of diffuse functions led us to select the 6-aug-cc-pVTZ basis set as the reference basis set to include the Gaussian continuum functions of Kaufmann {\it et al.}~\cite{kauf+89physb}. We have added 3, 5, and 8 Gaussian continuum functions (denoted by K) 
for each angular momentum in the 6-aug-cc-pVTZ basis set. 
In the lower half of Table~\ref{angmom}, the number of total, bound, and continuum states and the maximum energy obtained with these 6-aug-cc-pVTZ+3K, 6-aug-cc-pVTZ+5K, and 6-aug-cc-pVTZ+8K basis sets
is reported. It is noteworthy that increasing the number of K functions hardly affects the number of bound states, in favor of positive energy states, thus focusing the improvement on the description of the continuum. 
More precisely, as the maximum energy obtained with these three basis sets is nearly unchanged (6.313, 6.681, and 6.927 hartree, respectively), the K functions increase the density of states
in the energetically important region of the continuum.

\begin{figure}
\begin{center}
\includegraphics[width=3in,angle=0,trim=00mm 00mm 00mm 00mm,clip=true]{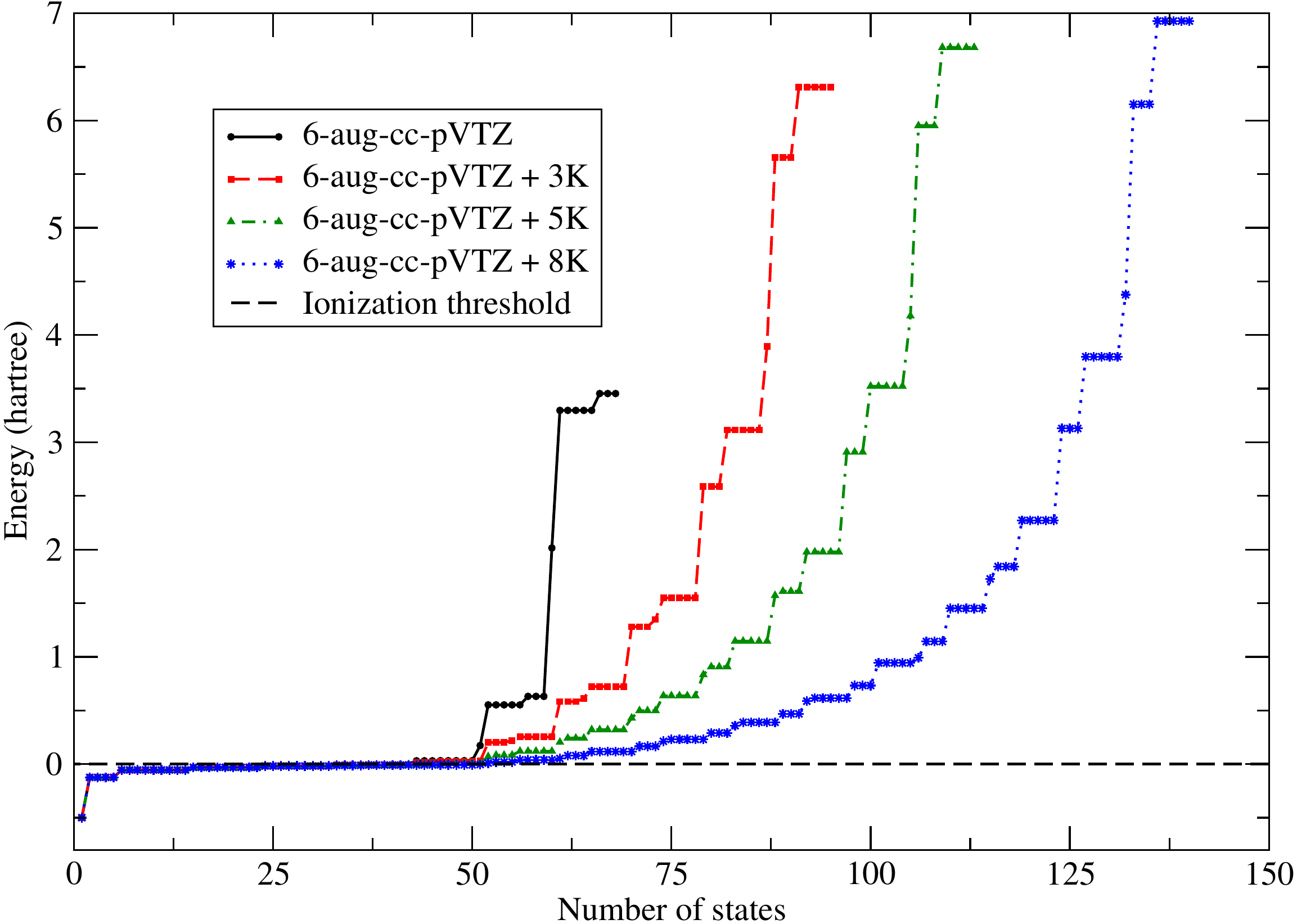}
\caption{Distribution of the state energies obtained with the 6-aug-cc-pVTZ basis set and increasing numbers of Gaussian continuum functions (K). \label{fig4}} 
\end{center}
\end{figure}

We show in Figure~\ref{fig4} the distribution of the state energies for the different basis sets. Increasing the number of K functions essentially does not change the energy spectrum
below the ionization threshold, while an almost continuum distribution builds up in the low-energy region above the ionization threshold. When compared with the 6-aug-cc-pVTZ basis set, 
the distribution of the continuum states becomes more dense (closer to a ``real'' continuum) and the gaps between (near-)degenerate sets of states become smaller. In particular, 
the density of states is improved in the region from the ionization threshold to around 1 hartree, which is also the most relevant energy region for HHG for the laser
intensity range studied here, according to the three-step model. 
By contrast, Luppi and Head-Gordon \cite{lupp+13jcp} showed that adding diffuse functions increases the density of Rydberg states, leaving the density of continuum states mostly unchanged.

\begin{figure}[t]
\begin{center}
\includegraphics[width=3in,angle=0]{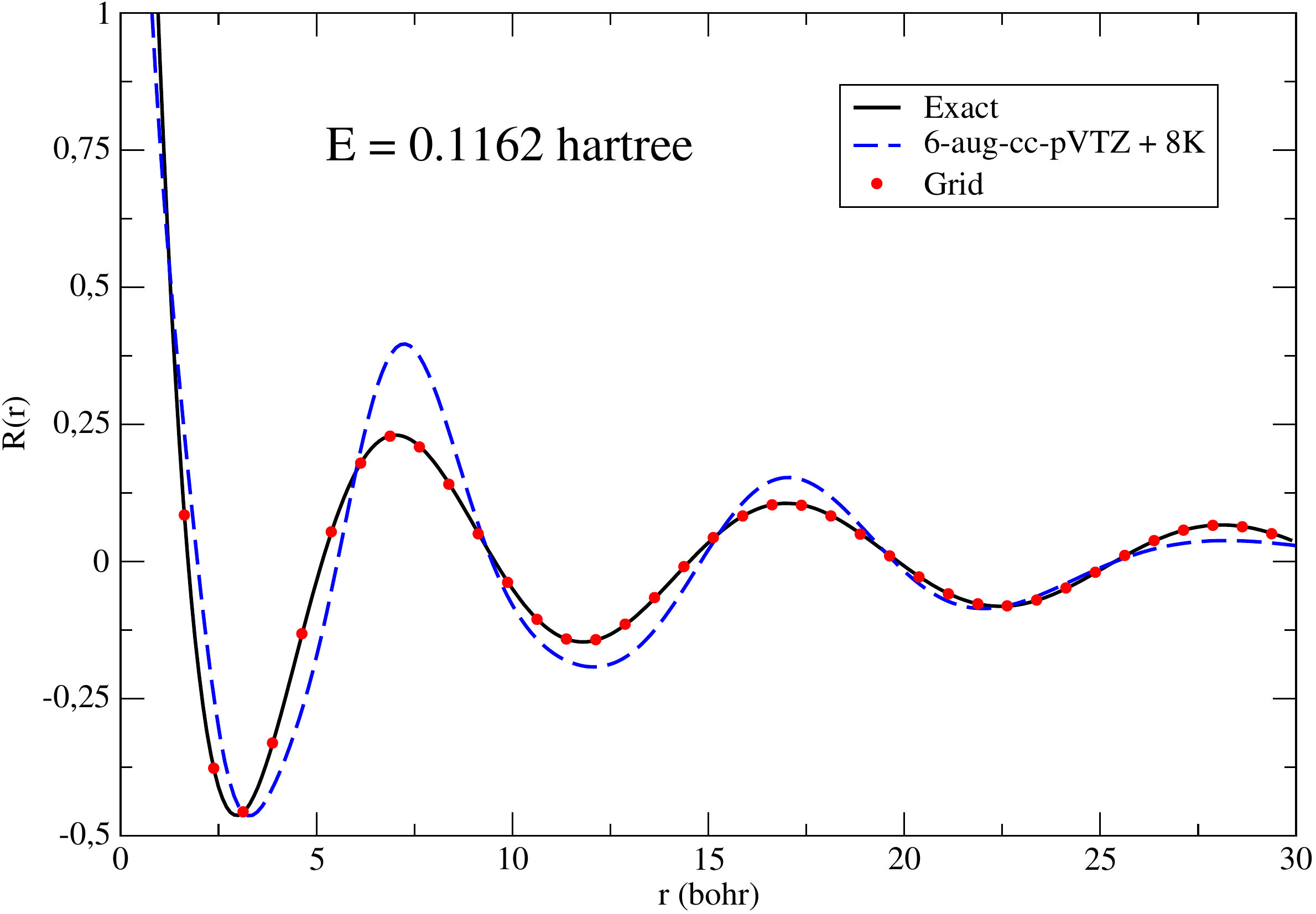} \\
\includegraphics[width=3in,angle=0]{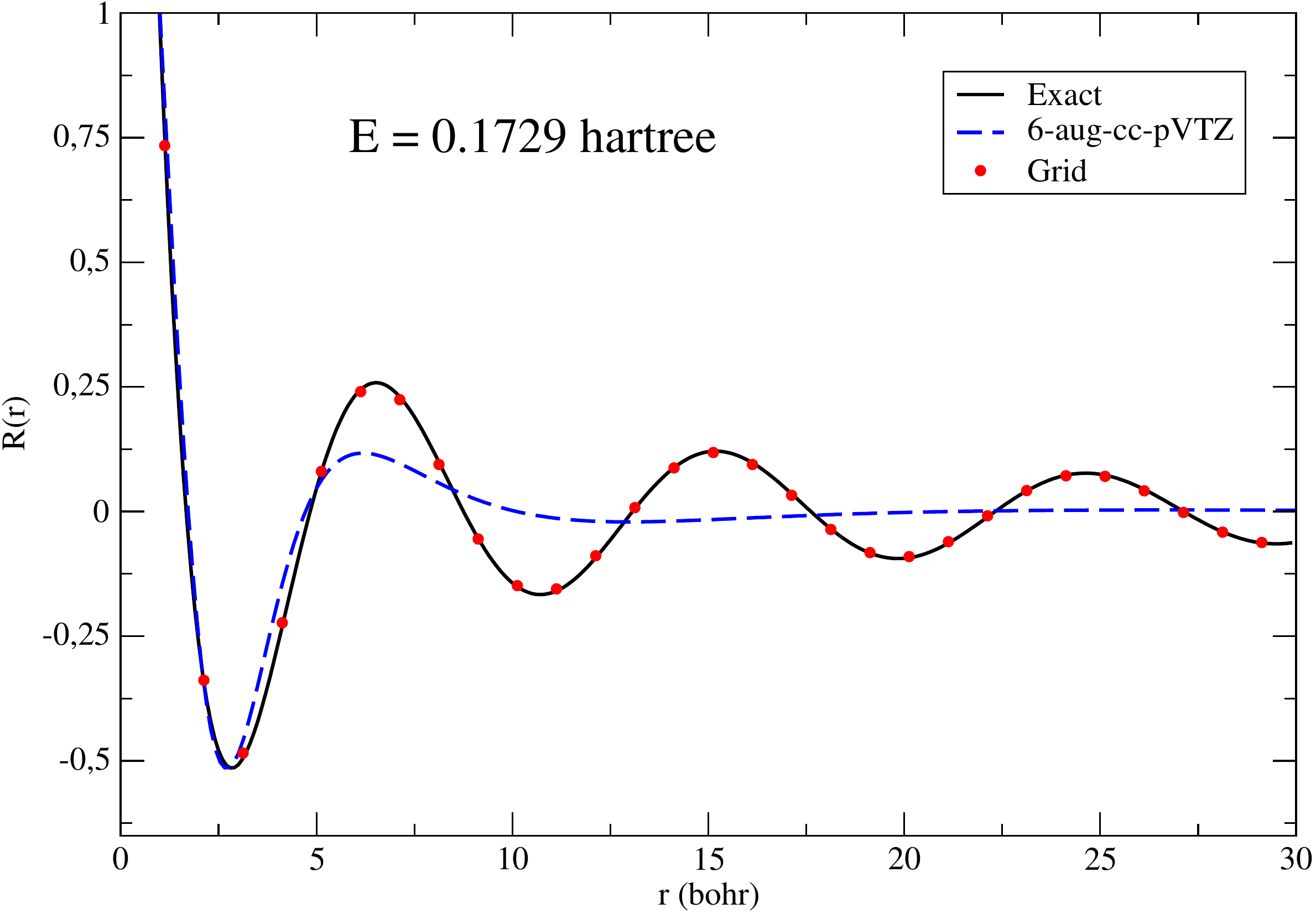}
\caption{Comparison between the exact radial wave function $R(r)$~\cite{BetSal-BOOK-57} and the radial wave function obtained using the 6-aug-cc-pVTZ+8K basis set for a s continuum state at the energy $E=0.1162$ hartree (upper panel). In the lower panel, the same comparison is done for a similar state of close energy $E=0.1729$ hartree but with the 6-aug-cc-pVTZ basis set, i.e. without the Kaufmann basis functions. The radial wave functions obtained in the grid calculations are also shown. Since continuum wave functions cannot be normalized in the standard way, the curves have been scaled in order to approximately have the same value at the first minimum.}
\label{fig6}
\end{center}
\end{figure}

The upper panel of Figure~\ref{fig6} compares the radial wave function $R(r)$ of a s continuum state at the energy $E=0.1162$ hartree obtained with the 6-aug-cc-pVTZ+8K basis set
 with the analytical solution of the time-independent Schr{\"o}dinger equation~\cite{BetSal-BOOK-57}. For completeness, the radial wave function from the grid calculation is also shown and is perfectly superimposed with the analytical solution. 
The radial wave function obtained with the 6-aug-cc-pVTZ+8K basis set is a reasonable approximation to the exact solution, the continuum Gaussian functions correctly reproducing the oscillations
of the function up to a radial distance as large as 30 bohr. This radial distance is consistent with the maximum distance $R_{\text{max}}$ (see Table \ref{tab:lasers}) 
traveled by the electron predicted by the three-step model with the laser parameters used here. 
For comparison, the lower panel of Figure~\ref{fig6} shows the radial wave function $R(r)$ obtained with the 6-aug-cc-pVTZ basis set for a similar s continuum state at the closest energy
 obtained with this basis set, $E=0.1729$ hartree. Clearly, the basis set without the continuum Gaussian functions is only able to describe the short-range part of function $R(r)$ but
 not the long-range oscillating part.

\begin{figure}
\begin{center}
\includegraphics[width=3in,angle=0]{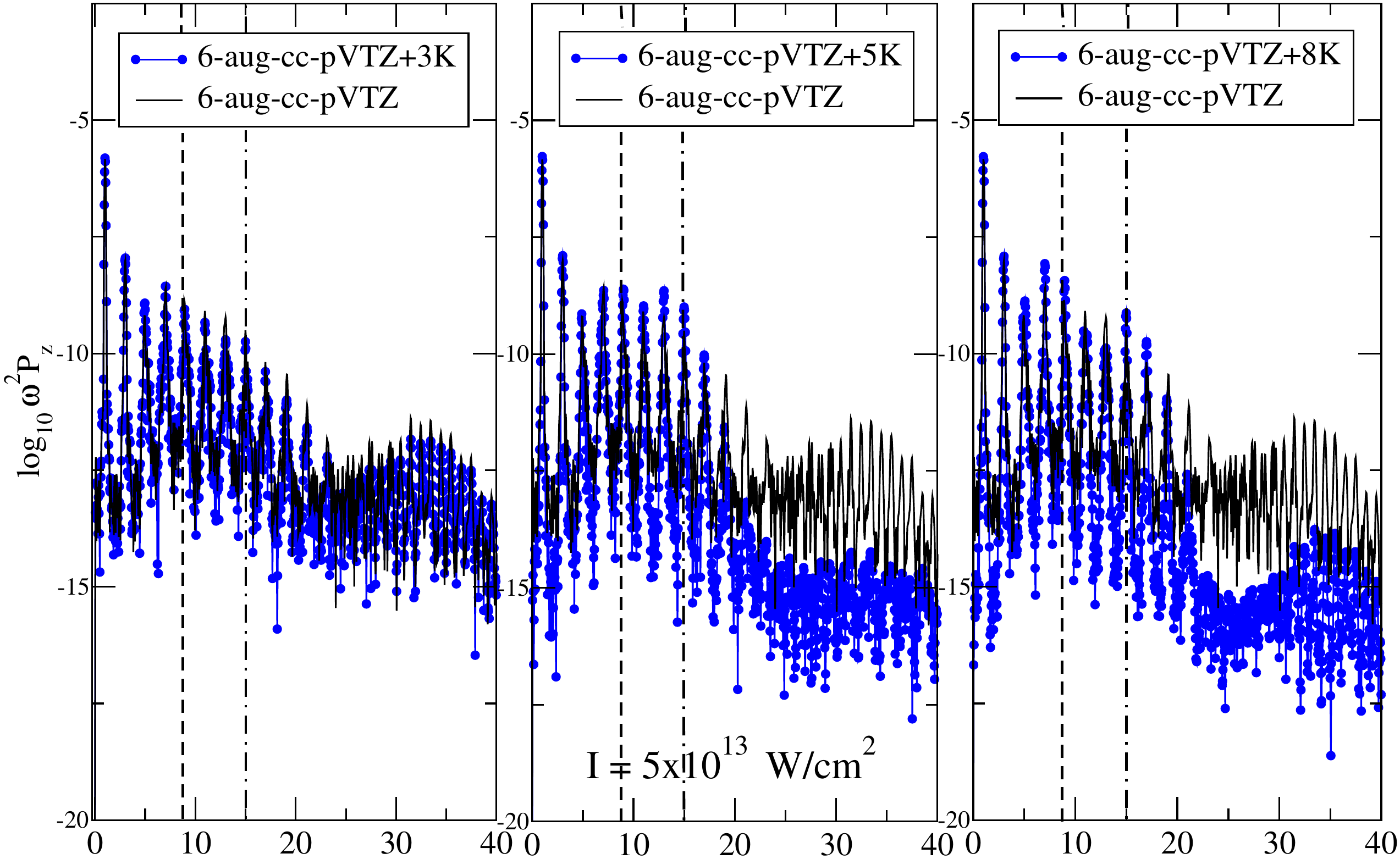} \\
\includegraphics[width=3in,angle=0]{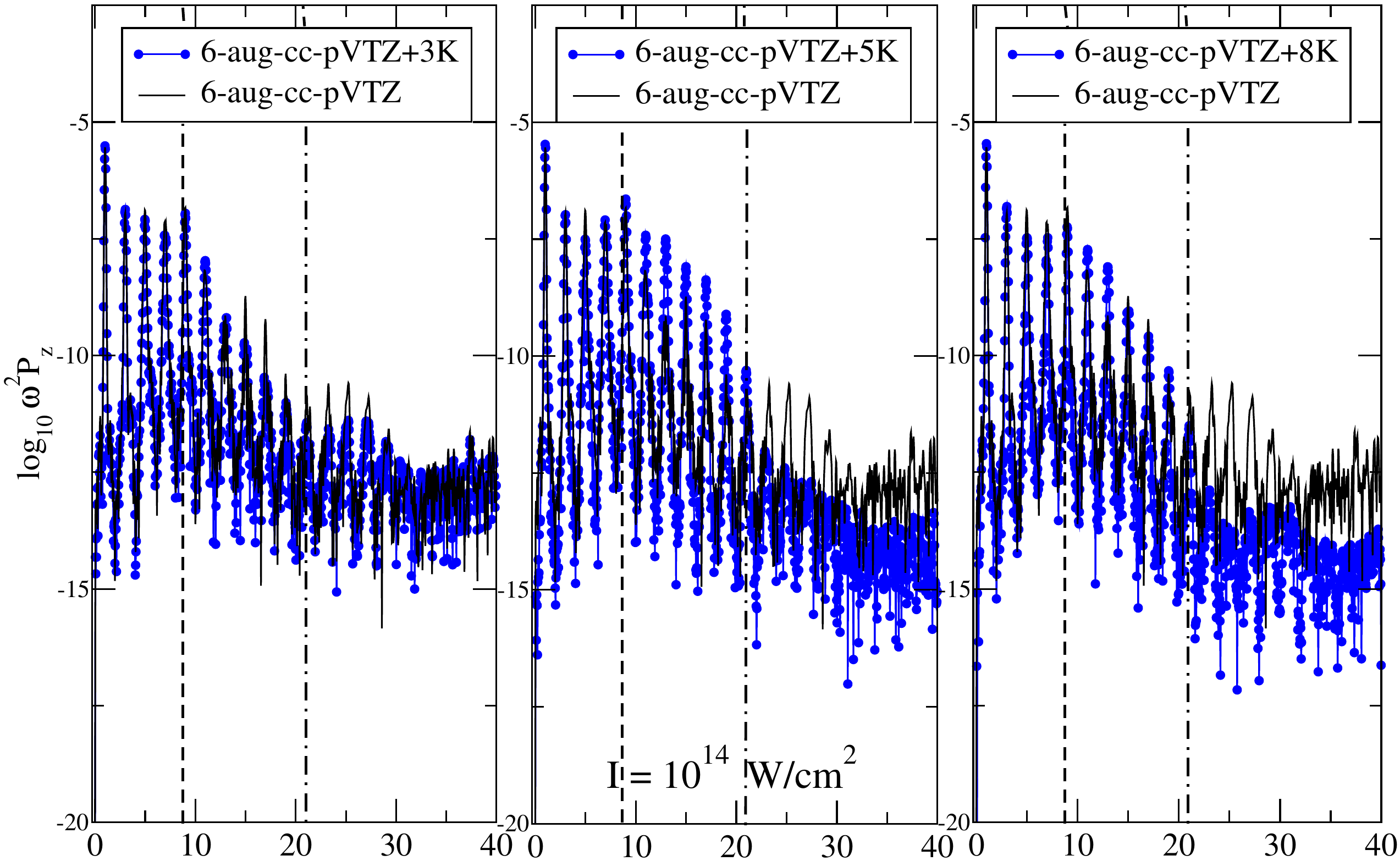}   \\
\includegraphics[width=3in,angle=0]{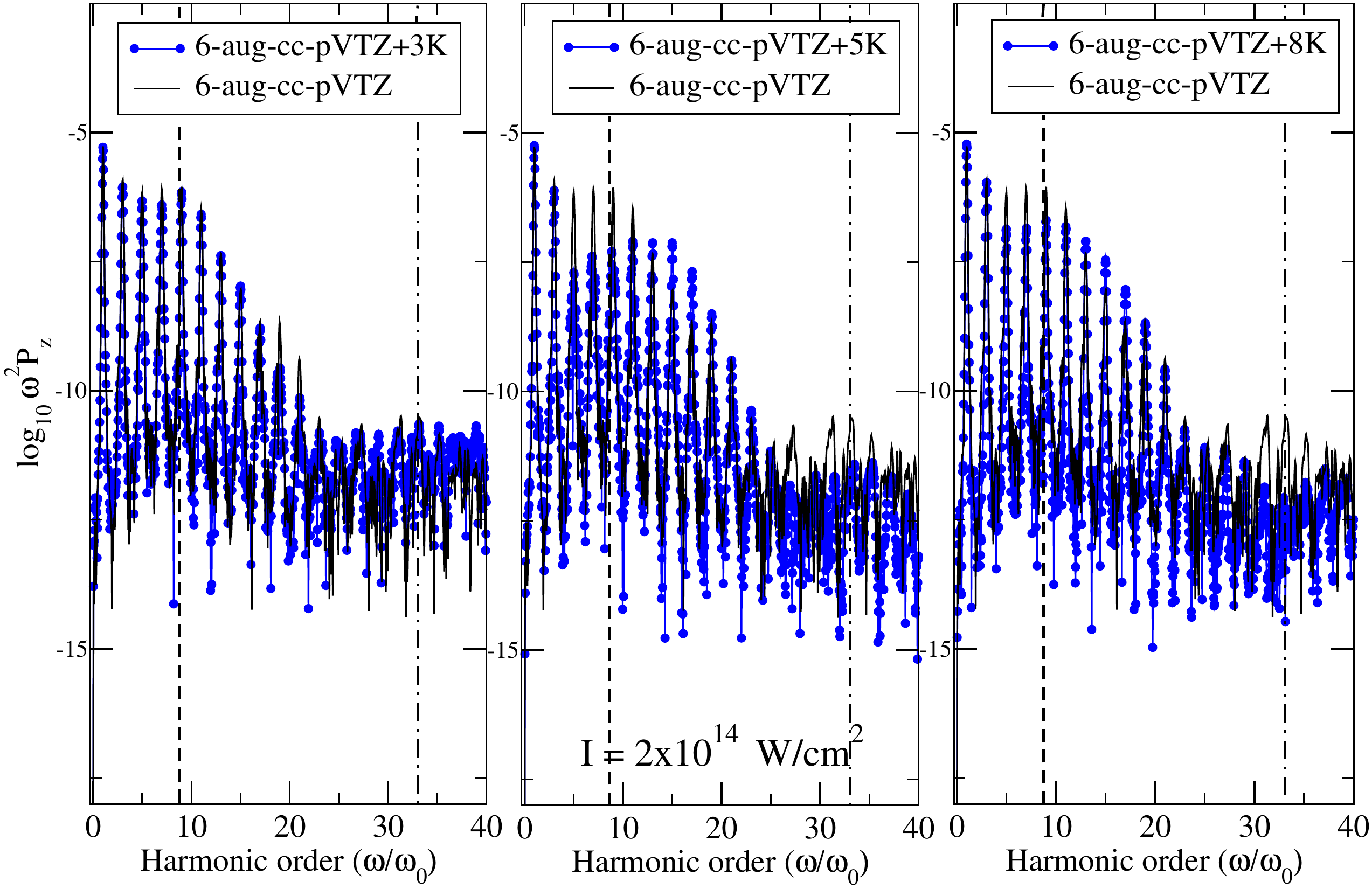}
\caption{Comparison among the velocity HHG spectra of the H atom extracted from the dipole power spectrum $\omega^2 P_z(\omega)$ calculated with the 6-aug-cc-pVTZ basis set plus 3 (left), 5 (middle), and 
8 (right) Gaussian continuum (K) functions. The laser intensity is $I = 5 \times$10$^{13}$ W/cm$^{2}$ (top), $I=10^{14}$ W/cm$^{2}$ (middle), and $2\times$10$^{14}$ W/cm$^{2}$ (bottom). 
The ionization threshold ($I_\text{p}/\omega_0$, vertical dashed line) and the harmonic cutoff in the three-step model $N_{\text{cutoff}}$ (vertical dot-dashed line) are also shown.
\label{fig5}} 
\end{center}
\end{figure} 

In Figure~\ref{fig5} the velocity HHG spectrum extracted from the dipole power spectrum is shown for the 6-aug-cc-pVTZ+3K, 6-aug-cc-pVTZ+5K, and 6-aug-cc-pVTZ+8K basis sets and for 
the three laser intensities. We focus our attention to the post-cutoff background region of the spectrum since diminishing the background in this region is an important goal of the present work.
Considering the laser intensity $I = 5\times 10^{13}$ W/cm$^2$ and analyzing the spectra between the 20th and 40th harmonics, we observe that the HHG spectrum with the 6-aug-cc-pVTZ+3K basis set
resembles the one obtained with the original 6-aug-cc-pVTZ basis set, with no obvious improvement. 
When adding 5 or 8 K functions the background is strongly diminished, while the harmonics before
the cutoff are not substantially changed. The same trend is also observed for laser intensities $I = 10^{14}$ W/cm$^2$ and $I = 2\times 10^{14}$ W/cm$^2$, even if the lowering of the background 
is not as strong. 

As demonstrated in Ref.~\citenum{lupp+13jcp}, the Rydberg bound states strongly contribute to the background of the HHG spectrum. The addition of Gaussian continuum functions to the basis set allows
 one to appropriately describe the low-lying continuum states, leading to a more balanced basis set yielding a lower background and therefore a much clearer
identification of the cutoff region. Of course, such an improvement depends on the intensity of the laser pulse, since larger intensities require to describe continuum states of higher energy and therefore require more Gaussian continuum functions.

\subsection{Comparison with grid calculations and improvement of the lifetime model}

\begin{figure}
\begin{center}
\includegraphics[width=3in,angle=0]{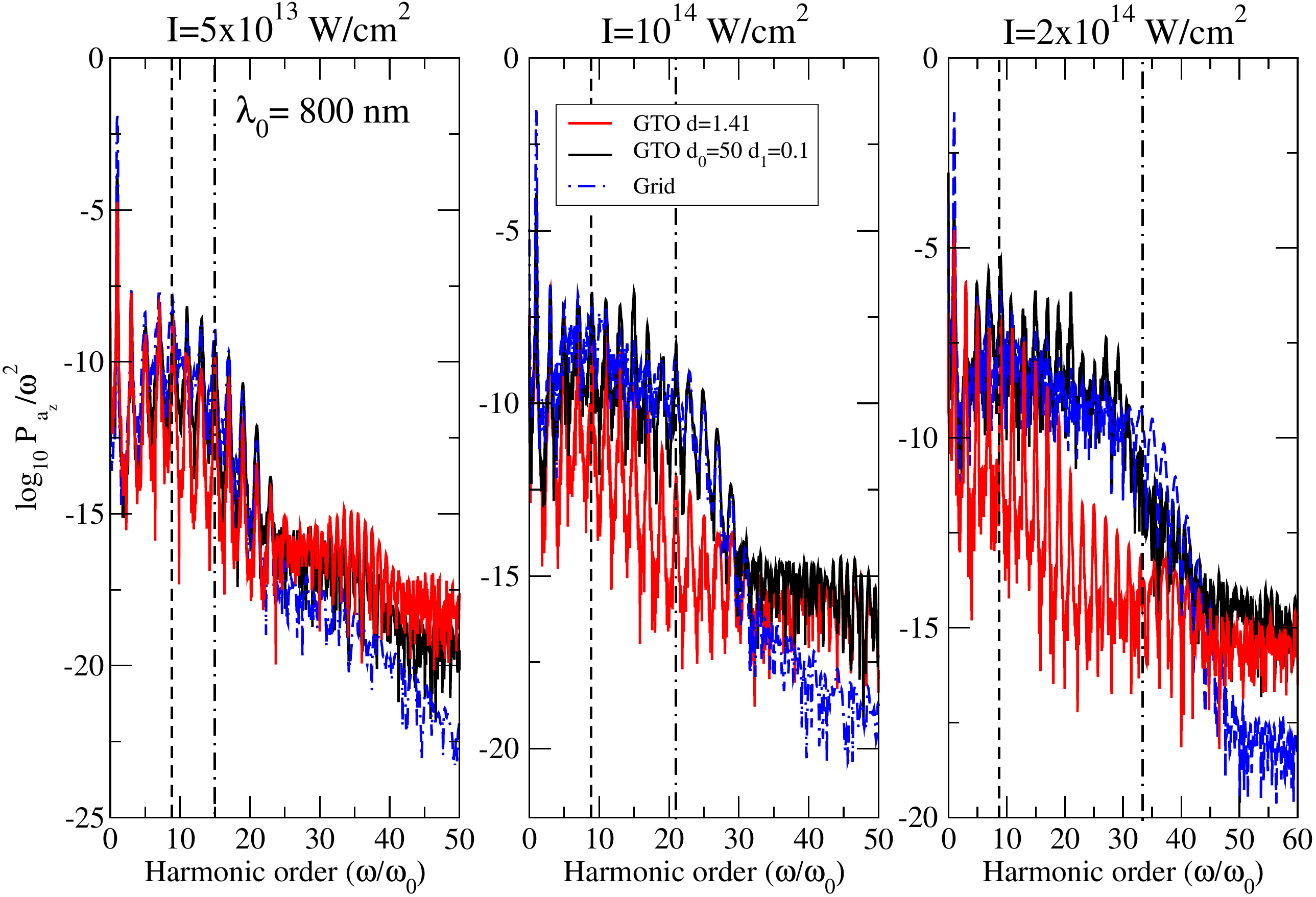}
\includegraphics[width=3in,angle=0]{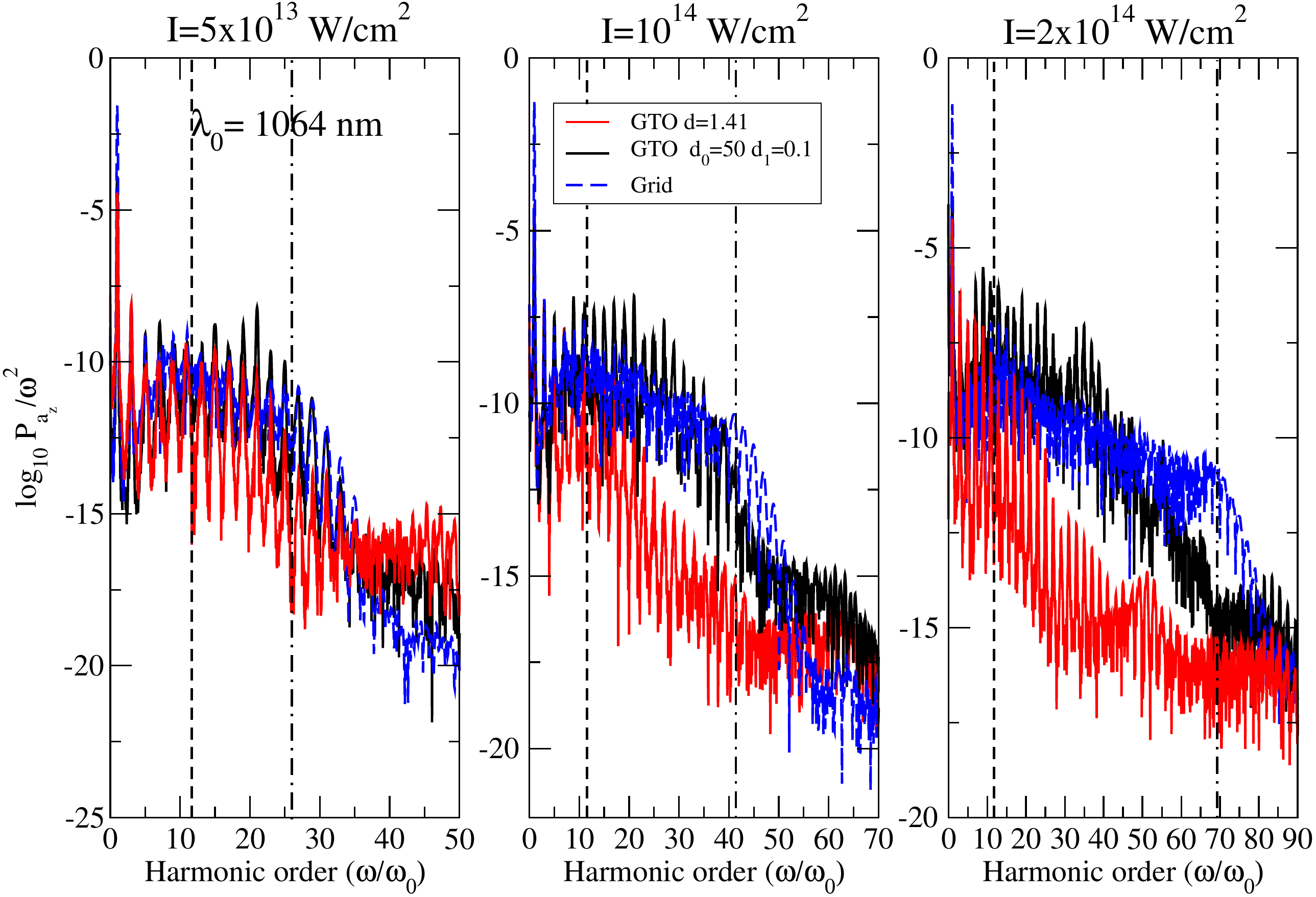} 
\caption{Velocity HHG spectrum extracted from the acceleration power spectrum $P_{a_z}(\omega)/\omega^2$ calculated with the 6-aug-cc-pVTZ+8K basis set with two lifetime models and with grid calculations, 
for the two laser wavelength $\lambda_0=800$ nm (upper panel) and 1064 nm (lower panel), the laser intensities $I=5 \times$10$^{13}$, 10$^{14}$, and 2$\times$10$^{14}$ W/cm$^2$.
The ionization threshold ($I_\text{p}/\omega_0$, vertical dashed line) and the harmonic cutoff in the three-step model $N_{\text{cutoff}}$ (vertical dot-dashed line) are also shown.
\label{fig8}}
\end{center}
\end{figure}

\begin{figure}
\begin{center}
\includegraphics[width=3in,angle=0]{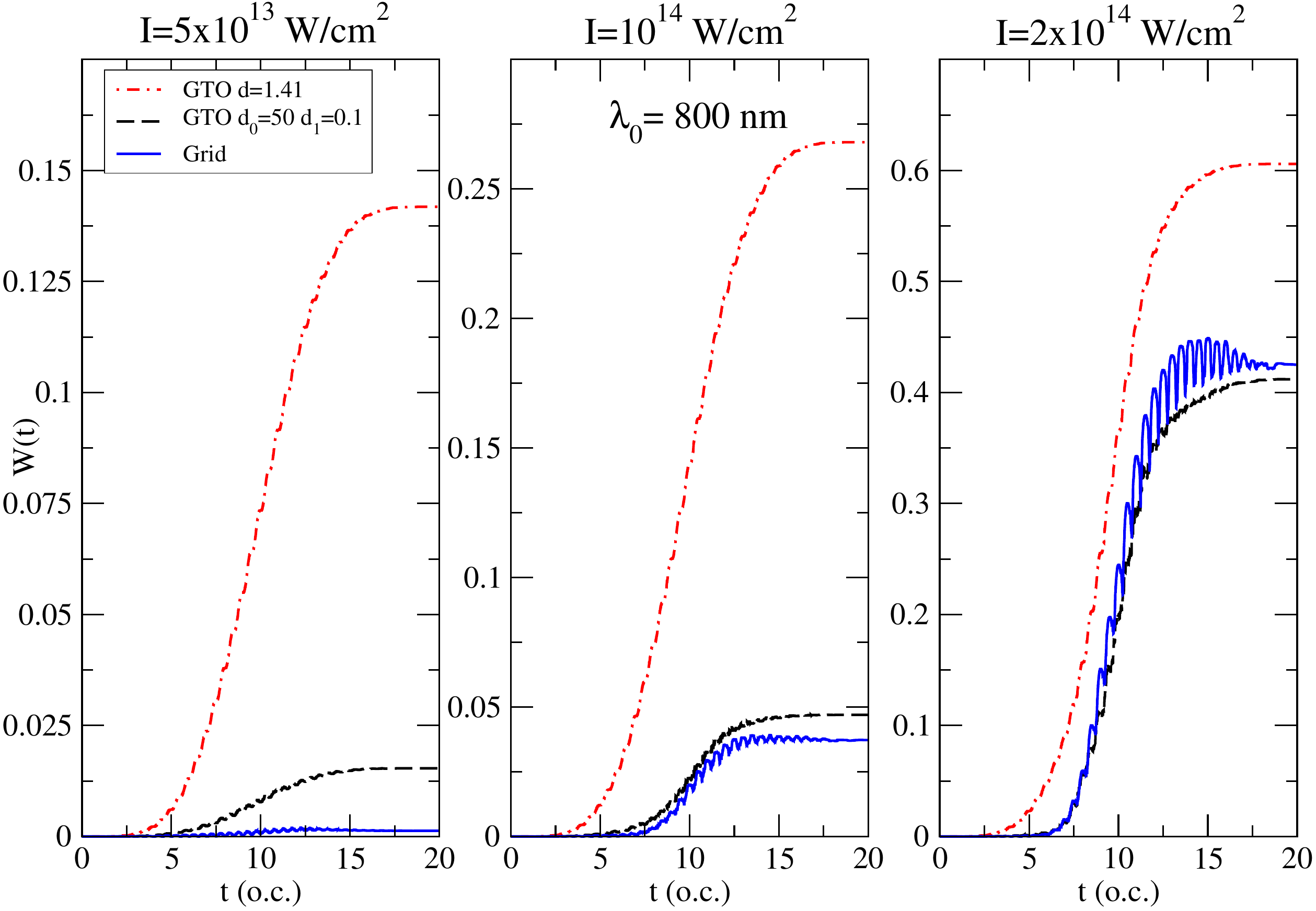} 
\caption{Ionization probability $W(t)$ for the laser wavelength $\lambda_0=800$ and the intensities $I = 5 \times$10$^{13}$ W/cm$^{2}$ (left), $I=10^{14}$ W/cm$^{2}$ (middle), and $2\times$10$^{14}$ W/cm$^{2}$ (right) obtained with the 6-aug-cc-pVTZ+8K basis set with the original (one parameter) and modified (two parameters) lifetime models, compared with the results from the grid calculations.}
\label{fig7}
\end{center}
\end{figure}

We now investigate in more detail the performance of the 6-aug-cc-pVTZ+8K basis set by comparison with grid calculations. 

In Figure~\ref{fig8} we compare the velocity HHG spectrum extracted from the acceleration power spectrum, obtained with the 6-aug-cc-pVTZ+8K basis set with two lifetime models, 
and with grid calculations for the same laser intensities as before and for $\lambda_0=800$ and 1064 nm.
The acceleration power spectrum, rather than the dipole power spectrum, was chosen here because the grid calculation is easier to converge for the acceleration power spectrum.
For the intensity $I=5 \times$10$^{13}$ W/cm$^2$ and for the two wavelengths, the spectra obtained with the Gaussian basis set and the original lifetime model (with $d=1.41$ bohr) are in good agreement with the ones from the grid calculations. In particular, the cutoff appears at almost the same energy. However, for the larger intensities, the intensity of the higher harmonics in the plateau obtained with the Gaussian basis set decrease too rapidly. This can be
attributed to a limitation of the original lifetime model which assigns too large lifetimes to high-energy continuum states. 

For this reason we introduce a modified version of the lifetime model, with two different values of the parameter $d$: a large value, $d_0=50$ bohr, 
for continuum states with positive energies below the energy cutoff of the three-step model $E_\text{cutoff}-I_\text{p}$, and a small value, $d_1=0.1$ bohr,
for continuum states with energies above $E_\text{cutoff}-I_\text{p}$. The ionization rates $\Gamma_k$ are thus smaller than in the original model for low-lying continuum states,
and larger for high-lying continuum states. This choice allows us to get a more accurate description of the harmonics in the plateau and close to the cutoff. 
The values of $d_0$ and $d_1$ have been chosen comparing with the corresponding grid HHG spectra. Not surprisingly, the value of $d_0$
 is of the order of magnitude of the electron excursion distance $R_{\text{max}}$ (see Table \ref{tab:lasers}).

We test our modified lifetime model by calculating the ionization probability (for both the grid and Gaussian-basis-set calculations)
\begin{equation}
W(t) = 1 - \sum_{k}^{\text{bound}} \vert \langle \psi_k \vert \Psi(t)  \rangle  \vert^2,
\end{equation}
where the sum runs over all the bound states. Figure~\ref{fig7} reports $W(t)$ obtained with the original and the modified lifetime models and from the grid calculations for the three laser intensities. The original lifetime model leads to largely overestimated ionization probabilities in comparison to the results obtained from the grid calculations. Our modified lifetime model reduces the ionization probability and is in better agreement with the grid calculations, especially for the intensities $I = 10^{14}$ W/cm$^2$ and $I = 2\times 10^{14}$ W/cm$^2$.

Coming back to Figure~\ref{fig8}, it is seen that the combined use of the 6-aug-cc-pVTZ+8K basis set and of the modified lifetime model results in a HHG spectrum that is in good agreement with the one obtained with the grid calculation at wavelength $\lambda_0=800$ nm and the intensity $I=10^{14}$ W/cm$^2$. The general shape of the spectrum and the position of the harmonic cutoff are well reproduced with the Gaussian basis set, the only remaining differences being larger
peaks and a larger background after the cutoff in comparison to the grid results. For the same wavelength and the largest intensity $I=2\times$10$^{14}$ W/cm$^2$, the agreement is also fairly good even though the position of the harmonic cutoff predicted with the Gaussian basis set is slightly too low.

The longer laser wavelength $\lambda_0=1064$ nm represents a more stringent test for our method since higher-energy regions are probed (see Table \ref{tab:lasers}). 
The agreement between the HHG spectra obtained with the Gaussian basis set and from the grid calculations is still pretty good for the intensity $I=5 \times$10$^{13}$ W/cm$^2$, 
while the position of the cutoff is slightly underestimated for the intensity $I=10^{14}$ W/cm$^2$ and significantly underestimated for the
largest intensity $I=2\times 10^{14}$ W/cm$^2$. This likely comes from a too poor description of the continuum states above 1 hartree with the 6-aug-cc-pVTZ+8K basis set, 
which can be populated for these wavelengths and intensities. A larger number of continuum Gaussian functions is needed in order to improve the high-energy part of the HHG spectrum 
for the largest intensities. We note, however, that increasing the number of continuum Gaussian functions can lead to near-linear dependencies in
the basis set (seen with the presence of very small eigenvalues of the overlap matrix of the basis functions) and thus numerical instability issues in self-consistent-field calculations.

\section{Conclusions}
\label{conclusions}

In this work, we have explored the calculation of the velocity HHG spectrum of the H atom extracted from the dipole, velocity, and acceleration power spectra with Gaussian basis sets for 
different laser intensities and wavelengths. 
While all the three power spectra give reasonable velocity HHG spectra with similar harmonic peaks before the cutoff, they tend to differ in the background region beyond the cutoff. 
The HHG spectrum extracted from the dipole power spectrum is the most sensitive to the basis set. With the 6-aug-cc-pVTZ basis set it leads to a high background which blurs out the location of the plateau cutoff.

Increasing the cardinal number of the basis set (from $X=\text{T}$ to $X=5$) or the number of diffuse basis functions (from $N=6$ to $N=9$) does not improve the HHG spectrum.
By contrast, adding 5 or 8 Gaussian continuum functions, as proposed by Kaufmann {\it et al.}~\cite{kauf+89physb}, leads to an improvement of the velocity HHG spectrum extracted from the dipole power spectrum at least 
for laser intensities up to $10^{14}$ W/cm$^2$ by decreasing the background, which thus allows one to better identify the cutoff region. 

The combined use of Gaussian continuum functions and a heuristic lifetime model with two parameters for modeling ionization results is in a fairly good agreement with the reference HHG spectra 
from grid calculations, in terms of the general shape of the spectrum, the number and intensity of peaks, and the position of the cutoff. 
The agreement is less satisfactory for the largest intensities because the high-energy continuum states are poorly reproduced by 
the Gaussian basis set calculations. 
Improving the accuracy for the largest intensities would require a larger number of Gaussian continuum functions.

Gaussian continuum functions thus appear as a promising way of constructing Gaussian basis sets for studying electron dynamics in strong laser fields, allowing one to define a balanced basis set 
to properly describe both bound and continuum eigenstates. The present work therefore opens the way to the systematic application of well established quantum chemistry methods with Gaussian basis sets to the study of highly nonlinear phenomena (such as HHG, photoionization cross sections, above-threshold ionization rates,...) in atoms and molecules.

\section*{Acknowledgements}
This work was supported by the Labex MiChem and CalSimLab part of French state funds managed by the ANR within the Investissements d'Avenir programme under references ANR-11-IDEX-0004-02 and ANR-11-LABX-0037-01. We thank A. Savin for having pointed out to us Ref.~\onlinecite{kauf+89physb} and V. V\'eniard for useful comments.
     
\appendix
\section{Relationship between the dipole, velocity, and acceleration forms of the power spectrum}
\label{app:powerspectrum}

In this appendix, we review the relationship between the dipole, velocity, and acceleration forms of the power spectrum~\cite{Burnett:1992, Bandrauk:2009ig, Han10}. If we define $\xi(t) = \bra \Psi(t) \vert \hat{\xi} \vert \Psi(t) \ket$, where $\xi$ stands for position $z$, velocity $v_z$, or acceleration $a_z$, and its Fourier transform
\begin{equation}
\xi(\omega)=\int^{t_\f}_{t_\i} \xi(t) e^{-i \omega t} \d t,
\label{xiomega}
\end{equation}   
the three forms of the power spectrum are commonly expressed as
\begin{equation}
P_\xi (\omega) = \frac{1}{(t_\f -t_\i)^2} |\xi(\omega)|^2.
\label{Pxi}
\end{equation}   
and the relationship between the three forms is the relationship between the three $|\xi(\omega)|^2$.

Applying Eq.~(\ref{xiomega}) for $\xi=v_z$, performing an integration by parts over $t$, and using $v_z(t) = \d z(t)/\d t$, gives
\begin{eqnarray}
{v_z}(\omega) = {z}(t_\f) e^{-i \omega t_\f}  -  {z}(t_\i) e^{-i \omega t_\i} + i \omega {z}(\omega), 
\end{eqnarray}     
which, if we have the condition ${z}(t_\i)= 0$, can be simplified as
\begin{equation}
{v_z}(\omega)= {z}(t_\f) e^{-i \omega t_\f} + i \omega {z}(\omega).
\end{equation} 
The relation between $|{z}(\omega)|^2$ and $|{v_z}(\omega)|^2$ is then
\begin{eqnarray}
|{v_z}(\omega)|^2 = \omega^2 |{z}(\omega)|^2 &+& \Bigl( {z}(t_\f)^2 
- 2\; \omega {z}(t_\f) \text{Im} [ {z}(\omega)e^{i \omega t_\f}]\Bigl), \;
\nonumber\\
\label{vel}
\end{eqnarray} 
which, in the case where we can make the approximation ${z}(t_\f) \approx 0$, simplifies as 
\begin{equation}
|{v_z}(\omega)|^2 \approx \omega^2 |{z}(\omega)|^2.
\label{vel_rel}
\end{equation} 
Similarly, applying now Eq.~(\ref{xiomega}) for $\xi=a_z$, using $a_z(t) = \d v_z(t)/\d t$, and integrating by parts gives
\begin{equation}
{a_z}(\omega)= {v_z}(t_\f) e^{-i \omega t_\f}  + i \omega {v_z}(\omega),
\label{acc}
\end{equation}  
where we have used the condition ${v_z}(t_\i)= 0$. This leads to the relation between $|{v_z}(\omega)|^2$ and $|{a_z}(\omega)|^2$
\begin{eqnarray}
|{a_z}(\omega)|^2 = \omega^2 |{v_z}(\omega)|^2 &+& \Bigl( {v_z}(t_\f)^2 
- 2\; \omega {v_z}(t_\f) \text{Im} [ {v_z}(\omega)e^{i \omega t_\f}]\Bigl),
\nonumber\\
\end{eqnarray} 
which, if we can make the approximation ${v_z}(t_\f) \approx 0$, gives the final approximate relations between the three forms of the spectrum
\begin{equation}
\omega^2 P_{z}(\omega) \approx P_{v_z}(\omega) \approx \frac{1}{\omega^2} P_{a_z}(\omega).
\label{spectra}
\end{equation}

\end{document}